%% file: main.tex
\documentclass[sigconf]{acmart}
\usepackage{subfig}
\usepackage{tcolorbox}
\usepackage{tikz}
\usepackage{multirow}
\usepackage{soul}
\usepackage{pgfplots}
\usepackage{ifthen}
\usepackage{xifthen}
\usepackage{hyperref}
\definecolor{taskColor}{HTML}{c2e9fb} 
\definecolor{guideColor}{HTML}{f3ccd4}
\definecolor{descriptionColor}{HTML}{ebedee}
\definecolor{formatColor}{HTML}{F7DD99}
\AtBeginDocument{%
  \providecommand\BibTeX{{%
    \normalfont B\kern-0.5em{\scshape i\kern-0.25em b}\kern-0.8em\TeX}}}

\setcopyright{acmlicensed}
\copyrightyear{2024}
\acmYear{2024}
\acmDOI{XXXXXXX.XXXXXXX}

\acmISBN{978-1-4503-XXXX-X/18/06}

\pgfplotsset{compat=1.16,
    tick label style={
        font=\tiny,
        /pgf/number format/sci,
    },
}

\newcommand{\overviewbarplot}[4]
{\begin{tikzpicture}
    \begin{axis} [
        title={#1},
        title style={font=\bfseries\tiny,at={(0.5,0.85)}},
        ybar=0.5pt, height=3.5cm, width=5cm, bar width=0.7pt,
        xtick={1,2,3,4,5,6,7,8,9,10,11,12,13,14,15,16,17,18,19,20},
        y tick label style={/pgf/number format/.cd,std,precision=2},
        xticklabels={#2},
        x tick label style={rotate=270,anchor=west,/pgf/number format/1000 sep=},
        scaled ticks=true,
        enlargelimits=0.05,
        ylabel style={font=\tiny},
        legend style={font=\tiny, legend pos=outer north east,legend columns=-1},
        axis lines*=left,
        tick align=outside,
        ymajorgrids=true,
    ]
    \addplot [draw=red, fill=red] coordinates {#3};
    \addplot [draw=blue, fill=blue] coordinates {#4};
    \end{axis}
\end{tikzpicture}}

\newcommand{\compbarplot}[4]
{\begin{tikzpicture}
    \begin{axis} [
        title={#1},
        title style={font=\bfseries\tiny,at={(0.5,0.8)}},
        ybar=2pt, height=3.5cm, width=5cm, bar width=8pt,
        xtick={1,2,3},
        y tick label style={/pgf/number format/.cd,std,precision=2},
        xticklabels={Precision, Recall, F1},
        x tick label style={/pgf/number format/1000 sep=},
        scaled ticks=true,
        enlargelimits=0.15,
        ylabel style={font=\tiny},
        legend style={font=\tiny, at={(1,-0.25)},legend columns=-1},
        axis lines*=left,
        tick align=outside,
        ymajorgrids=true,
    ]
    \addplot [draw=red, fill=red] coordinates {#2};
    \addplot [draw=blue, fill=blue] coordinates {#3};
    \legend{#4}
    \end{axis}
\end{tikzpicture}}

\newcommand{\largetemplot}[1]
{\begin{tikzpicture}
    \begin{axis} [
        height=4cm, width=7cm,
        xtick={0.1,0.2,0.3,0.4,0.5,0.6,0.7,0.8,0.9,1.0},
        y tick label style={/pgf/number format/.cd,std,precision=2},
        x tick label style={/pgf/number format/.cd,std,precision=2},
        scaled ticks=true,
        xlabel={Temperature},
        xlabel style={font=\tiny, at={(0.5,-0.1)}},
        ylabel={F1 score},
        ylabel style={font=\tiny, at={(-0.1,0.5)}},
        legend style={font=\tiny, legend pos=outer north east,legend columns=1},
    ]
    {#1}
    \end{axis}
\end{tikzpicture}}

\newcommand{\templotWithLabel}[4]
{\begin{tikzpicture}
    \begin{axis} [
        title={#1},
        title style={font=\bfseries\tiny,at={(0.5,0.95)}},
        height=4cm, width=5cm,
        xtick={0.1,0.2,0.3,0.4,0.5,0.6,0.7,0.8,0.9,1.0},
        y tick label style={/pgf/number format/.cd,std,precision=2},
        x tick label style={/pgf/number format/.cd,std,precision=2},
        scaled ticks=true,
        xlabel={#2},
        xlabel style={font=\tiny, at={(0.5,-0.1)}},
        ylabel={#3},
        ylabel style={font=\tiny, at={(-0.1,0.5)}},
        legend style={font=\tiny, legend pos=outer north east,legend columns=-1},
    ]
    {#4}
    \end{axis}
\end{tikzpicture}}

\newcommand{\addlineWithLegend}[3]{
\addplot[mark=#2] plot coordinates {#3};
\addlegendentry{#1}
}

\newcommand{\addline}[3]{
\addplot[mark=#2] plot coordinates {#3};
}

\begin{document}

\title[LLM-Base Domain Modeling With Question Decomposition]{A Model Is Not Built By A Single Prompt:\\ LLM-Based Domain Modeling With Question Decomposition}

\author{Ru Chen}
\orcid{0009-0000-9960-9534} % TODO: get an orcid number
\email{chenru@ustb.edu.cn} % TODO: fill your emails
\affiliation{%
  \institution{School of Computer and Communication Engineering, University of Science and Technology Beijing}
  \streetaddress{No. 30, Xueyuan Road, Haidian district}
  \city{Beijing}
  \country{China}
  \postcode{100083}
}

\author{Jingwei Shen}
\orcid{0009-0009-6380-4456} % TODO: get an orcid number
\email{jingweishen2022@126.com} % TODO: fill your emails
\affiliation{%
  \institution{School of Computer and Communication Engineering, University of Science and Technology Beijing}
  \streetaddress{No. 30, Xueyuan Road, Haidian district}
  \city{Beijing}
  \country{China}
  \postcode{100083}
}

\author{Xiao He}
\orcid{0000-0002-3000-0795}
\authornote{Corresponding author.}
\email{hexiao@ustb.edu.cn}
\affiliation{%
  \institution{School of Computer and Communication Engineering, University of Science and Technology Beijing}
  \streetaddress{No. 30, Xueyuan Road, Haidian district}
  \city{Beijing}
  \country{China}
  \postcode{100083}
}

\newtheorem{remark}{Remark}


\begin{abstract}
Domain modeling, a crucial part of model-driven engineering, demands extensive domain knowledge and experience from engineers.
When the system description is highly complicated, the modeling task can become particularly challenging and time-consuming. 
Large language Models(LLMs) can assist by automatically generating an initial object model from the system description.
Although LLMs have demonstrated remarkable code-generation ability, they still struggle with model-generation using a single prompt. 
In real-world domain modeling, engineers usually decompose complex tasks into easily solvable sub-tasks, significantly controlling complexity and enhancing model quality.
Inspired by this, we propose an LLM-based domain modeling approach via question decomposition, similar to developer's modeling process. 
Following conventional modeling guidelines, we divide the model generation task into several sub-tasks, i.e., class generation, association and aggregation generation, and inheritance generation.
For each sub-task, we carefully design the prompt by choosing more efficient query words and providing essential modeling knowledge to unlock the modeling potential of LLMs.
To sum up all the sub-tasks solutions, we implemente a proof-of-object tool integrated into the standard Ecore editor that asks LLMs to generate an object model from the system description.
We evaluate our approach with 20 systems from different application domains.
The preliminary results show that our approach outperforms the single-prompt-based prompt by improving recall values and F1 scores in most systems for modeling the classes, attributes, and relationships.
% We expect our work to aid development of tools and techniques, especially LLM-based, designed for domain modeling.
% Our tool is publicly available at [link]，and demo video can be found at [link]. 
\end{abstract}

\begin{CCSXML}
<ccs2012>
   <concept>
       <concept_id>10011007.10010940.10010971.10010980.10010984</concept_id>
       <concept_desc>Software and its engineering~Model-driven software engineering</concept_desc>
       <concept_significance>500</concept_significance>
       </concept>
   <concept>
       <concept_id>10010147.10010178.10010179</concept_id>
       <concept_desc>Computing methodologies~Natural language processing</concept_desc>
       <concept_significance>300</concept_significance>
       </concept>
 </ccs2012>
\end{CCSXML}

\ccsdesc[500]{Software and its engineering~Model-driven software engineering}
\ccsdesc[300]{Computing methodologies~Natural language processing}

\keywords{Model generation, Large language models, Prompt engineering, Question decomposition}


\maketitle

\input{section/intro}
\input{section/relatedwork}
\input{section/approach}
% \input{section/implementation}
\input{section/evaluation}

\input{section/threats}
\input{section/conclusion}

% \begin{acks}
% \end{acks}


\balance

\bibliographystyle{ACM-Reference-Format}
\bibliography{mainref}

% \appendix


\end{document}

%% file: section/intro.tex
\section{Introduction}\label{sec:intro}
% P1: LLM for SE: a) introduction to LLM (1-2 sentences); b) list major applications in SE (1-2 sentences); c) narrow down to code gen (1-2 sentences).
Large language models (LLM), such as GPT \cite{Achiam2023GPT4TR,Radford2019LanguageMA,10.1145/3368089.3409680}, LLaMA \cite{touvron2023llamaopenefficientfoundation,rozière2024codellamaopenfoundation}, PaLM \cite{chowdhery2022palmscalinglanguagemodeling}, and Claude \cite{TheC3}, are artificial intelligence (AI) programs, trained on huge textual data, supporting text comprehension, translation, and generation. 
LLMs have attracted a significant amount of attention in software engineering community \cite{software-engineering-community}. 
They are being widely explored and experimented with for various development tasks, including code generation \cite{chen2021evaluating,OpenAICodeX,cai24,nijkamp2023codegen,verilog-code-gen,white2023chatgpt,döderlein2023piloting}, comment generation \cite{LLM-SQL2Text-gen,commentGeneration}, test case generation \cite{kang2023large,gu2023llm,LLM-genTest,feng2024prompting}, bug localization and fixing \cite{peng2024domain,LLM-fix-bugs}, and code merge \cite{svyatkovskiy2022program}.

% P2: AI in MDE: 1) is a new direction and helpness; 2) previous research; 
In addition to code-related tasks, LLMs are also anticipated to streamline system abstraction and modeling processes, alleviating modelers' workload, enhancing efficiency, and bridging language barriers and domain knowledge gaps.
%LLM-assisted modeling is an emerging research topic in model-driven engineering (MDE).
%LLMs are expected to automate system modeling, reducing the burden on modelers, improving efficiency, and overcoming the challenges posed by language barriers or domain knowledge gaps.
Pioneering efforts \cite{Chen23,natural-language-Gen-UML,herwanto,Apvrille2024SystemAA} have explored the potential of LLM-aided model generation from textual specifications.
The basic idea is to query LLMs with a \textit{prompt}, comprising a task statement, a system description, and some output guidelines, to obtain a system model in a certain format.
% The preliminary results are encouraging, shedding light on this direction.
The preliminary results are shedding light on this direction.
%For example, Chen et al. \cite{Chen23} demonstrated that GPTs can produce a object model based on a system description.

% P3: Why LLM-based model generation is hard--identify the challenges in model generation using LLM：1）skills; 2) fewer corpora; 
Nevertheless, current research efforts have revealed the fact that LLM-based model generation is much more difficult than LLM-based code generation. 
We think that the major cause of the difficulty is twofold.
First, modeling requires a diverse set of skills, such as analytical and design capabilities, abstract thinking, and generalization abilities, which are not fully encompassed by LLMs.
Second, compared with source code, there are significantly fewer corpora related to modeling in the real world.
Consequently, LLMs are not adequately trained to effectively handle models.
% Due to the limited availability of requirement documents, we did not use fine-tuning but instead opted for prompt engineering methods.

% p4: question decomposition: 1) generalizability, and inspired from what (1-2 sentences); 2) its advantage (2 sentences/points); 
% In real-world domain modeling, engineers typically decompose complex tasks into easily manageable sub-tasks.
% Inspired by the divide-and-conquer strategy, question decomposition to solve complex LLM-based tasks in software engineering is a general solution\cite{question-decomposition/self-collaboration,question-decomposition/self-planning}.
% Nonetheless, determining how to effectively decompose these tasks remains a problem-specific and tricky.

% P5:our key insight
% 'advance the field by...'provides a stronger emphasis on the contribution?
In this paper, we aim to go one step further by proposing an LLM-based approach to the generation of object-oriented domain models utilizing question decomposition.
Our key insight is that \textit{model generation is such a complex question that cannot be solved by a single prompt}---it must be divided into several sub-questions, each of which concentrates on a specific aspect of the overall question.
By forcing LLMs to answer simpler sub-questions, we can improve the quality of generated models.
% In fact, the question decomposition is a general divide-and-conquer strategy, which is frequently used in real-world modeling.

% The benefits of this approach include: 1) Simplifying the modeling process by breaking it down into smaller, more specific tasks. 2) Enhancing the precision and accuracy of LLMs responses through tailored sub-task prompts.

% P6: Our approach: 1)how to decompose: class anf relationship generation; 2) each prompt strategy, key points; 3) narrow gap between llm and practical modeling, we implement a prototype tool; 
Following real-world modeling guidelines \cite{UMLTextbook,sepa20}, we organize the process of domain model generation into two sequential steps, i.e. class generation and relationship generation.
For class generation, we adopt a two-turn conversion strategy that asks LLMs to first identify and then refine class and attribute definitions.
For relationship generation, we further split it into two independent tasks, i.e. the generation of inheritances and the generation of associations and aggregations.
To unlock the modeling potential of LLMs, we carefully design the prompt used in each step by providing \textit{essential modeling knowledge} and choosing \textit{more efficient query words}.
To avoid syntactic model errors, we parse the generated results of all the sub-tasks and programmatically create an object model.
% P7: implementation & contribution
We have implemented a prototype tool as a Python package and integrated it into the standard Ecore editor on Eclipse platform.
To evaluate our approach, we conducted experiments using a dataset containing 20 system descriptions, coming from various application domains.
% To evaluate our approach, we conduct an experiment based on the dataset including the descriptions of 20 systems from different application domains.
% We adopt a structural similarity metric to compare the textual object models generated by our approach with those produced by the approach proposed in \cite{Chen23}.
% The experimental results show that our approach improves the recall values and F1 scores in most systems compared with the baseline.
The experimental results show that our approach outperforms the baseline in most systems.
% We also notice that (inspiration by LLM) ...?

The contributions of our paper are summarized as follows:
\begin{itemize}
    \item We propose an LLM-based object-oriented domain modeling approach via question decomposition. We also explore how to inject modeling knowledge into prompts to improve answers returned by LLMs.
    \item We conduct an empirical evaluation based on 20 software systems to compare our approach with Chen et al's approach \cite{Chen23}. We also discover that different sub-questions in our approach are sensitive to different temperature settings.
    \item We implement a prototype tool on Eclipse platform to promote LLM-aided domain modeling.
\end{itemize}

The rest of this paper is organized as follows.
Section \ref{sec:relatedwork} discusses the related work.
Section \ref{sec:approach} introduces the technical details of our approach.
% Section \ref{sec:tool} briefly demonstrates our prototype tool.
Section \ref{sec:evaluation} presents the evaluation and comparison with the baseline.
The last section concludes the paper and future work.

%% file: section/relatedwork.tex
\section{Related Work}\label{sec:relatedwork}

\paragraph{\textbf{NLP-based model generation}}
During the past decades, natural language processing (NLP) has been applied in MDE to achieve requirement modeling \cite{zhao21}, UML model generation \cite{Harmain}, and model-to-code transformation \cite{Franc}.
Particularly, Yue et al. \cite{Yue15} proposed the \textit{Restricted Use Case Modeling} (RUCM), which defines some restrictive rules to constrain the use of natural language in use case specifications.
After parsing restricted English into syntax trees, a model transformation is adopted to convert use case specifications into UML/SysML models.
The major limitation is that users must follow the restrictive rules and cannot freely write the specifications.
Despite the inconvenience, their approach has been applied in many real-life MDE projects, demonstrating its good practicality.

% P3
% Chen et al.\cite{Chen23} have conducted a comparative study of conceptual model generation using LLMs. 
% They tried zero-shot, N-shot, and chain-of-thought (CoT) prompts on GPT 3.5 and GPT 4.0 with 10 systems from different application domains. 
% They compared the results and found that the combination of GPT 4.0 and 1-shot prompt achieved the best performance. However, they pointed out that there is still a significant gap between their LLM-based method and real-life applications, as the recall values and F1 scores of most
% systems tested are not satisfactory.

% 接下来就是讲文献work
% 现有研究中LLM的单步提示应用
% p4. 引入单步提示的普遍性：概述单步提示在LLM辅助建模任务中的普遍应用，强调其作为一种直接且简便的方法，如何迅速获得研究和开发社区的关注。 

% P3
Recent studies \cite{camera,Chen23,Chen23b,herwanto,Achiam2023GPT4TR} mainly focused on LLM-aided model generation with carefully designed prompts.
Most of them attempted to achieve model generation by using a single prompt.

\begin{figure}[!tb]
    \centering
    \includegraphics[width=\columnwidth]{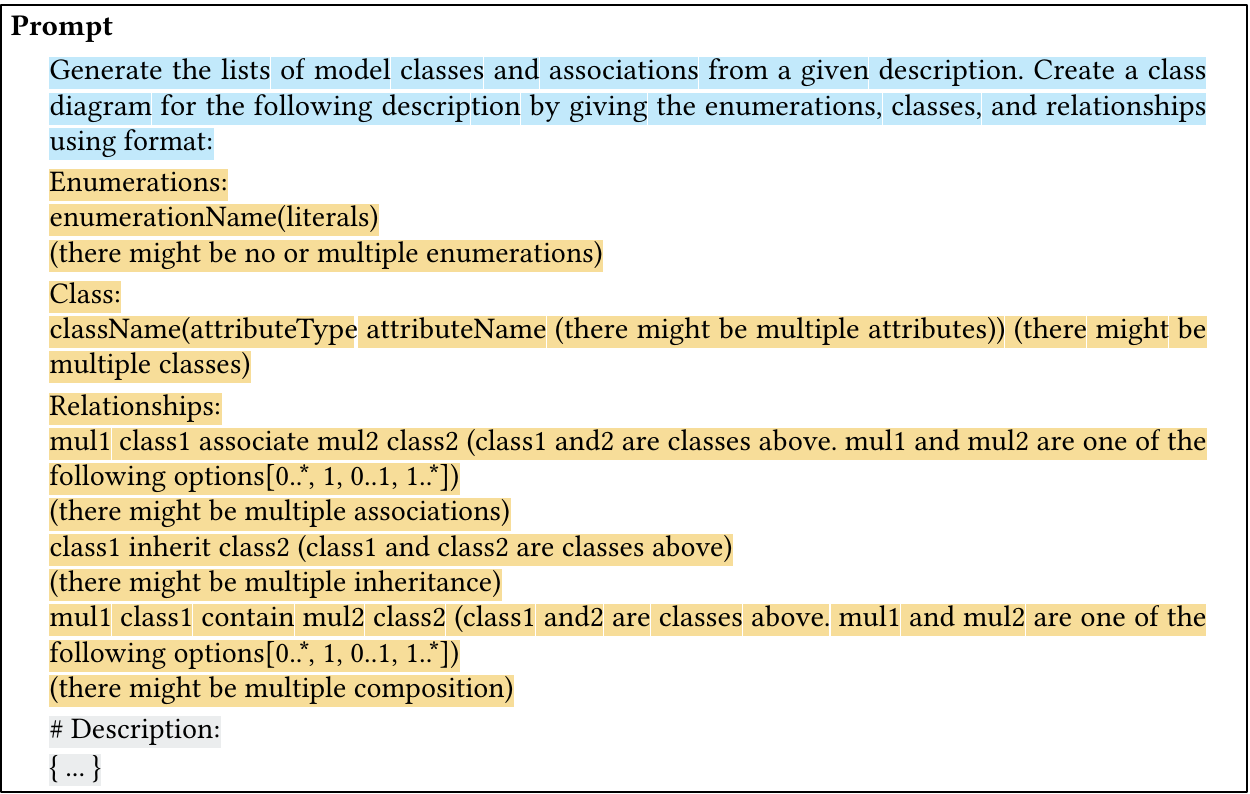}
    \caption{The zero-shot prompt from Chen et al. \cite{Chen23}}
    \label{fig:chen-zero-shot}
\end{figure}

Chen et al.\cite{Chen23} evaluated the effectiveness of different prompts for object-oriented domain model generation based on GPT 3.5 and GPT 4.0 using 10 systems.
They proposed and compared four types of prompts, including zero-shot, one-shot, two-shot, and chain-of-thought (CoT) prompts.
As shown in Figure \ref{fig:chen-zero-shot}, their zero-shot prompt consists of a task statement (in blue), a format description (in yellow), and a system description (in grey).
By filling in a concrete system description, a modeler can use this single prompt to ask LLMs to generate a domain model in a textual format.
Their N-shot prompts replace the format description with N examples, i.e., sample systems and the corresponding domain models.
Their CoT version requires LLMs to analyse a system description \textit{sentence by sentence} according to an illustrative example.
Further, Chen et al.\cite{Chen23b} explored the potential use of GPT 4 in goal-oriented modeling by comparing how the following information, including a single sentence describing basic requirements, contextual information, and syntax examples, affects model generation.

% p8
%  因此，我们受chen 方法的启发，做了本文工作。强调与chen方法的不同之处。
Our work is inspired by Chen et al. \cite{Chen23}, especially their zero-shot prompt as shown in Figure \ref{fig:chen-zero-shot}.
Nevertheless, our work is significantly distinct from \cite{Chen23} in the following two aspects.
(1) We decompose the process of domain model generation into several sub-questions.
LLMs shall answer each sub-question separately, rather than returning a model after a single interaction.
(2) We attempt to embed modeling guidelines and knowledge in our prompts to unlock the modeling capabilities of LLMs.

C\'{a}mara et al. \cite{camera} conducted an exploratory study on LLM-based
UML model generation. 
They investigated the capabilities of ChatGPT to generate PlantUML models and OCL expressions. 
The preliminary results showed that ChatGPT could generate various syntax and semantic model errors, and inconsistent responses. 
They pointed out that without human efforts, the current version of ChatGPT alone is insufficient to complete the modeling task, due to ChatGPT’s lack of abstraction capability.

% P5
% B. Chen et al. 使用GPT-4，通过一系列实验来评估GPT-4在目标建模方面的潜力和表现。% 他们设计并利用四种不同类型提示词来生成目标模型:
% 1) Single Sentence: 单一句子提示，描述了建模任务的基本要求；% 2) Single Sentence + Domain Paragraph: 在1）基础上提供了上下文信息；% 3) Syntax Description + Single Sentence: 在1）基础上增加了目标建模语句的语法描述；% 4) Syntax Description + Single Sentence + Domain Paragraph: 综合多种信息，提供GPT4.0最全面的提示信息。

% P4
% Mendez等人的工作：讨论Mendez等人如何使用单一提示在不同模型（GPT-3.5, GPT-4等）上实现数据流图的自动生成，突出他们方法的创新性以及在特定任务上的成功应用。 
Herwanto \cite{herwanto} demonstrated the feasibility of automating Data Flow Diagram generation using a single prompt on various LLMs, e.g. GPT-3.5, GPT-4, Llama2, and Mixtral models. Their prompt contains a task description, a detailed instruction, system requirements defined as user stories, and output templates.

% P6
% Apvrille et al.\cite{Apvrille2024SystemAA} 利用ChatGPT，使用单个提示词实现从系统规范生成SysML图形的自动化迭代过程。
% 他们构造了一个包含knowledge on system engineering、system specification、question type的提示词。
% 通过多次迭代 使得每次迭代的目的：更正、完善refine SysML code。
% 在迭代过程中，先前生成的sysml code作为反馈被重新引入到提示词中。
% 每次迭代是在已有成果的基础上进一步发展，逐渐完善SysML图形。
% Apvrille的迭代过程：强调他们如何通过单一提示实现从系统规范到SysML图形的自动化迭代过程。在这里，要特别指出Apvrille方法中迭代步骤的增加，这不仅是对单步提示的扩展，也是向更复杂建模任务迈进的一步。
Apvrille et al. \cite{Apvrille2024SystemAA} introduced an LLM-based approach to the generation of both structural and behavioral SysML models according to system specifications.
Different from other work, this approach is an \textit{iterative} approach, which involves multiple interactions with LLMs.
Each iteration will correct and refine the SysML models generated by the previous iteration.
It also checks the syntactic and semantic constraints against the newly generated model, and will send the error information to the next iteration.
Although our approach also requires multiple interactions with LLMs to generate a model, our approach is not iterative.

% Through multiple iterations, the purpose of each iteration is to correct and refine SysML code.
% During the iteration, the previously generated sysml code is reintroduced into the prompt as feedback.
% Each iteration is a further development on the basis of the existing results, gradually improve the SysML graph.

% P4 -> P7  这部分没改动

\paragraph{\textbf{LLMs in software engineering}}
% P1
Besides models, LLMs have already been used to generate other software artifacts \cite{LLM4SE-survey}, e.g. source code and test cases.
% P2
% For code generation, 
% % there are two main paradigms, i.e., fine-tuning LLMs and prompt engineering.
% By fine-tuning, an LLM can be adapted for a specific task.
% For example, CodeX \cite{OpenAICodeX} is a fine-tuned GPT language model for code generation based on 159 GB Python files.
% Cai et al. \cite{cai24} fine tuned CodeGen \cite{nijkamp2023codegen}, an LLM for code, to generate exception handling code.
% Thakur et al. \cite{verilog-code-gen} proposed \textit{VeriGen} for Verilog by fine-tuning CodeGen. 
White et al. \cite{white2023chatgpt} presented a prompt pattern catalog to solve the common problems caused by LLM-based software engineering automation.
D\"oderlein et al. \cite{döderlein2023piloting} investigated whether variations of LLMs' input parameters (e.g., prompt and LLMs' temperature) can have an impact on the quality of the generated code.
The problem of code generation may also be decomposed into many sub-tasks and be solved in multiple steps \cite{question-decomposition/self-planning, question-decomposition/self-collaboration}.
Kang \cite{kang2023large} uses an LLM to generate bug reproducing test cases from bug reports. 
AdbGPT \cite{feng2024prompting} can generate test cases to reproduce Android bugs. 
Sch\"afer et al. \cite{LLM-genTest} proposed TestPilot, an adaptive LLM-based unit test case generation tool for JavaScript.

% \textbf{Other application}
% Zhang et al. \cite{requirement-engineering} conducted a preliminary evaluation of ChatGPT’s zero-shot requirement retrieval performance on two requirements analysis tasks over four data sets. 
% Although these results are only preliminary, they provide optimism that
% LLMs can be used as a support for efficient and effective requirements engineering.
% C\^amara et al. \cite{LLM-SQL2Text-gen} proposed a method of generating comment for SQL queries with LLMs to help developers better understand SQL code. 

% Includeing code summary, code merge, requirements analysis

\paragraph{\textbf{Prompt engineering}}
% 1. What is prompt. (done) What is prompt engineSering.(done) What major techniques are there(done)
% 2. Few-shot prompt. (done)
% 3. Cot(done)
% 4. Least-to-most(done)
% 5. Question decomposition(done)
% 6. CoVe (done)
% 7. Voting
% p1
A prompt is an input to LLMs to help them recall what was learned during pre-training \cite{prompt4downstream,few-shotImpoveQualify}. 
Prompt engineering is a new discipline for developing and optimizing prompts to efficiently use LLMs. 
The main techniques of prompt engineering include few-shot prompting and chain-of-thought prompting. 
\textit{Few-shot} prompting is an in-context learning technique, where we provide a small set of examples in the prompt to steer LLMs toward better performance. 
\textit{Zero-shot} prompting provides no example.
CoT prompting \cite{wei2022chain} simulates a human’s thinking process that handles complicated problems in a step-by-step manner. 
It instructs LLMs to solve a problem by following some intermediate reasoning steps.

% P3
% Since LLMs are limited in reasoning about complex questions, the divide-and-conquer strategy is frequently used \cite{Divide-and-Conquer}.
% Chain-of-thought (CoT) prompting \cite{wei2022chain} simulates human’s thinking process that handles complicated problems in a step-by-step manner. 
% It instructs LLMs to solve a problem by following some intermediate reasoning steps.
% Zhou et al. \cite{zhou2023leasttomost} proposed the least-to-most prompting technique, which is an extension to CoT. 
% It queries LLMs to decompose the problem into sub-problems first, and then asks LLMs to solve each sub-problem in sequence.

% To mitigate the hallucination issues of LLMs, \textit{self-ask} \cite{press-etal-2023-measuring} was proposed.
% To answer the original question, \textit{self-ask} guides LLMs to generate some follow-up questions. Then, it utilizes LLMs and external search engines to address the follow-up questions. Finally, it asks LLMs to summarize the final answer to the original question.

%CoVe asks LLMs to generate some verification questions to verify whether the answer to the original question is correct.

%% file: section/approach.tex
\section{Our Approach}\label{sec:approach}
\subsection{Overview}
\begin{figure*}[!t]
\centering
  \includegraphics[width=0.9\textwidth]{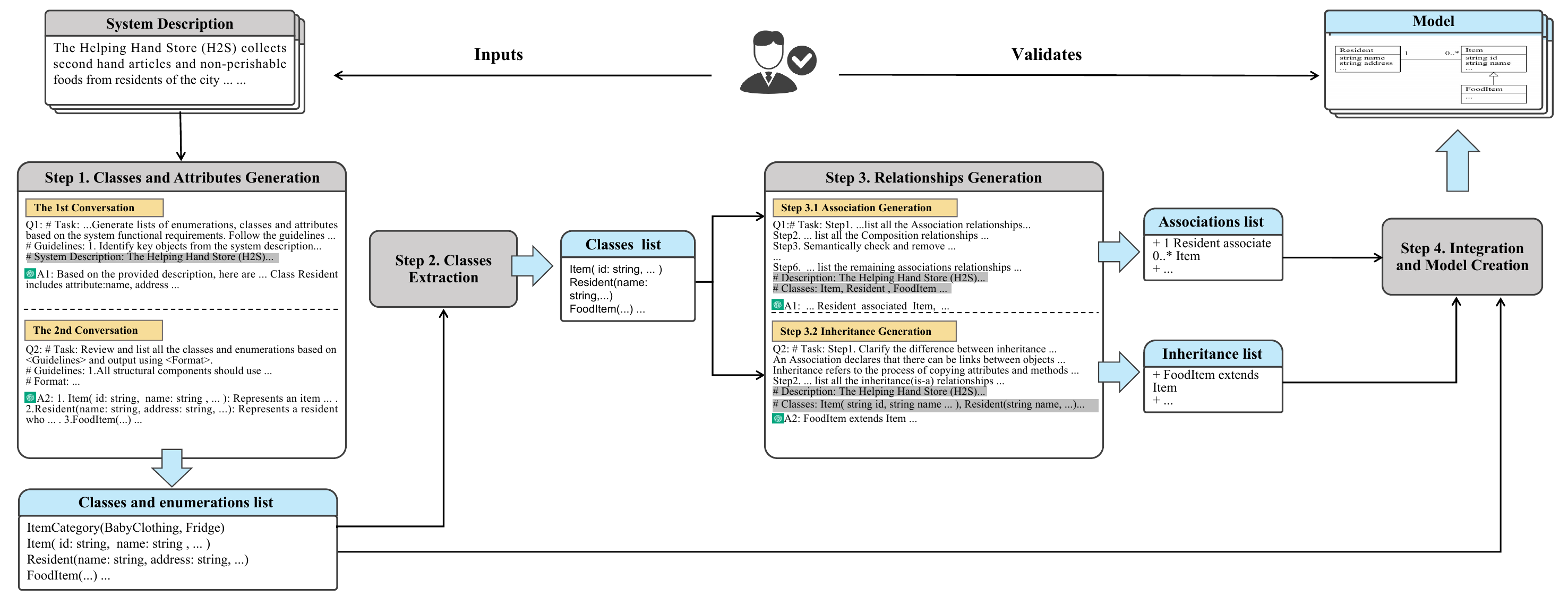}
  \caption{Approach overview}\label{fig:overview}
\end{figure*}

The general workflow of our approach is depicted in Figure \ref{fig:overview}.
The input of our approach is a textual system description, while the output is an object-oriented domain model created by an LLM according to the description.

Different from previous efforts that query an LLM with a single prompt, we assume that domain modeling is a complex question that currently cannot be solved by one-step generation.
The modeling guidelines and knowledge \cite{UMLTextbook,sepa20} for humans should be adopted to instruct an LLM in model generation.
Accordingly, our approach divides the task of model generation into four major steps.
\begin{itemize}
    \item \textbf{Step 1} takes the system description as input and identifies classes (as well as enumerations) and attributes mentioned in the description using a 2-turn conversation with an LLM (see Section \ref{sec:classgen}), resulting in definitions of classes, enumerations, and attributes.
    \item \textbf{Step 2} programmatically extracts out the class and attributes names by scanning the output of Step 1 with regular expressions.
    \item \textbf{Step 3} generates relationships by feeding both the system description and the classes into an LLM (see Section \ref{sec:relgen}). Our approach further splits Step 3 into two independent sub-steps, i.e., association generation (i.e. Step 3.1) and inheritance generation (i.e. Step 3.2).
    \item \textbf{Step 4} merges the outputs of the previous steps, parses them into structural data, and programmatically creates a class model (see Section \ref{sec:modelcreation}).
\end{itemize}

When applying our approach in practice, we believe that human modelers must be involved in the loop. 
Modelers must validate and refine the generated model to complete the modeling.

\subsection{Class generation}\label{sec:classgen}
%% NOTE: Be straightforward and insightful
%% NOTE: Keep in mind that our key point is question decomposition

The first step of our approach is \textit{class generation}. 
The input is the system description, while the output is the lists of classes and enumerations, as well as their attributes and enumeration literals.

In object-oriented modeling, identifying classes is a crucial task, having a profound impact on the subsequent steps. 
To improve the quality of class generation, we adopt a 2-turn conversation to list and refine the classes and enumerations. 
This way ensures that, during each turn, an LLM only needs to answer a simple question.

\begin{figure}[!t]
    \centering
    \subfloat[The first-round prompt\label{fig:class-gen-1st}]{\includegraphics[width=\columnwidth]{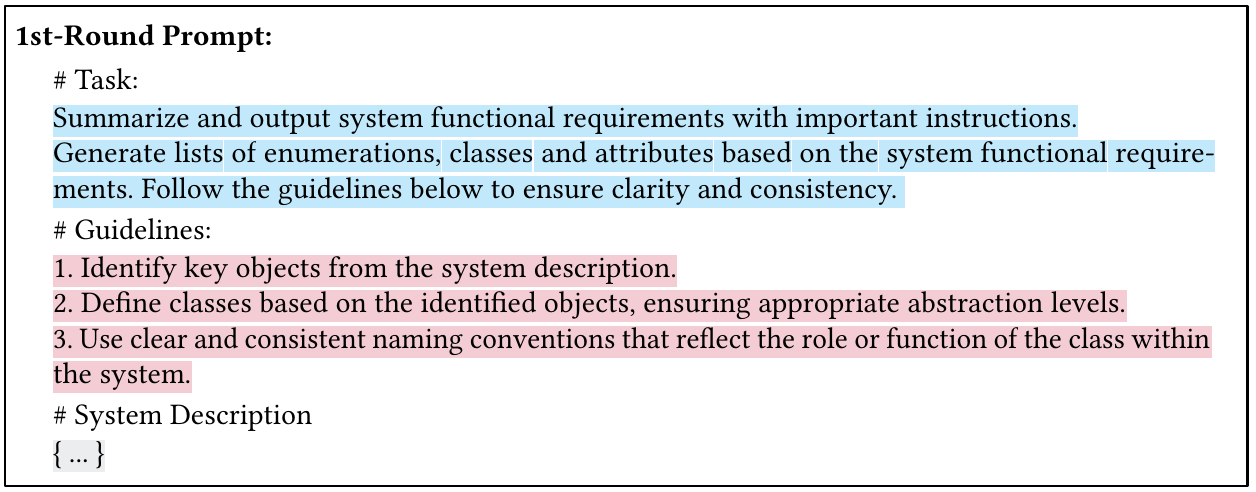}}\\
    \subfloat[The second-round prompt\label{fig:class-gen-2nd}]{\includegraphics[width=\columnwidth]{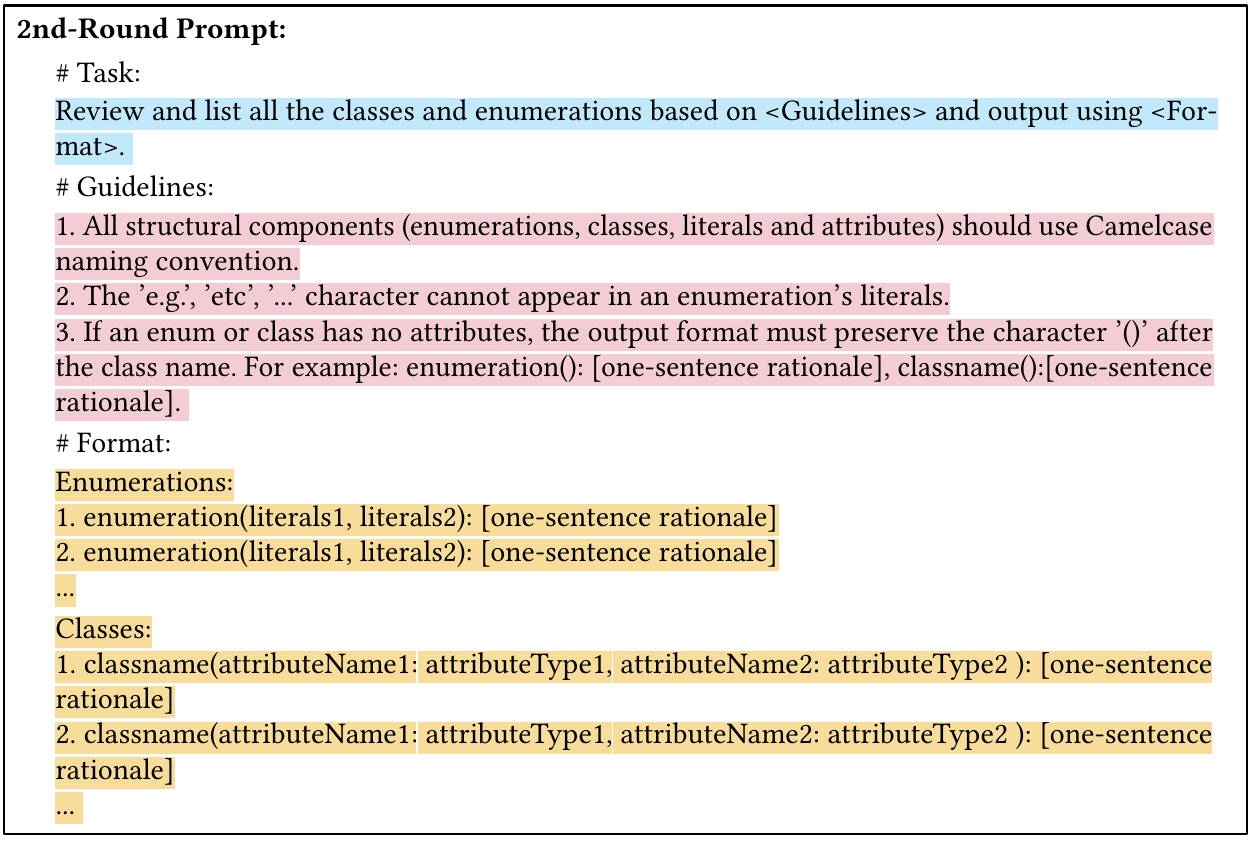}}
    \caption{Prompts of class generation (including \sethlcolor{taskColor}\hl{task statement}, \sethlcolor{formatColor}\hl{formatting rules}, \sethlcolor{guideColor}\hl{modeling knowledge}, and \sethlcolor{descriptionColor}\hl{system description})}\label{fig:class-gen}
\end{figure}

Figure \ref{fig:class-gen} illustrates the prompts for class generation.
The first-round prompt (see Figure \ref{fig:class-gen-1st}) consists of the task statement, the system description, and essential guidelines on object and class definition, requesting an LLM to generate enumerations, classes, and attributes according to the given description.
In the first round, an LLM concentrates on the task of analyzing the system description and is free to print out the answer in any format.

When an LLM returns an answer, we start the second round conversation, which takes both the system description and the first-round conversation as its context.
As shown in Figure \ref{fig:class-gen-2nd}, the second-round prompt requires an LLM to list again all the classes and enumerations using a specific format.
The format specifies that each class/enumeration, as well as its attributes/literals, is printed on one line to facilitate subsequent text parsing.
By learning from the contextual information, an LLM shall review and reformulate the first-round answer, having a chance to revise the output.

\subsection{Relationship generation}\label{sec:relgen}
After class generation, our approach will collect the definitions of classes and enumerations returned from Step 1.
Afterwards, we extract the generated classes (i.e. Step 2) using regular expressions.
These classes, combined with the system description, will be sent to Step 3---relationship generation, to produce the lists of \textit{inheritance}, \textit{association}, and \textit{aggregation} relationships.

\begin{figure}[!t]
    \subfloat[The prompt for associations and aggregations\label{fig:rel-gen-assoc}]{\includegraphics[width=\columnwidth]{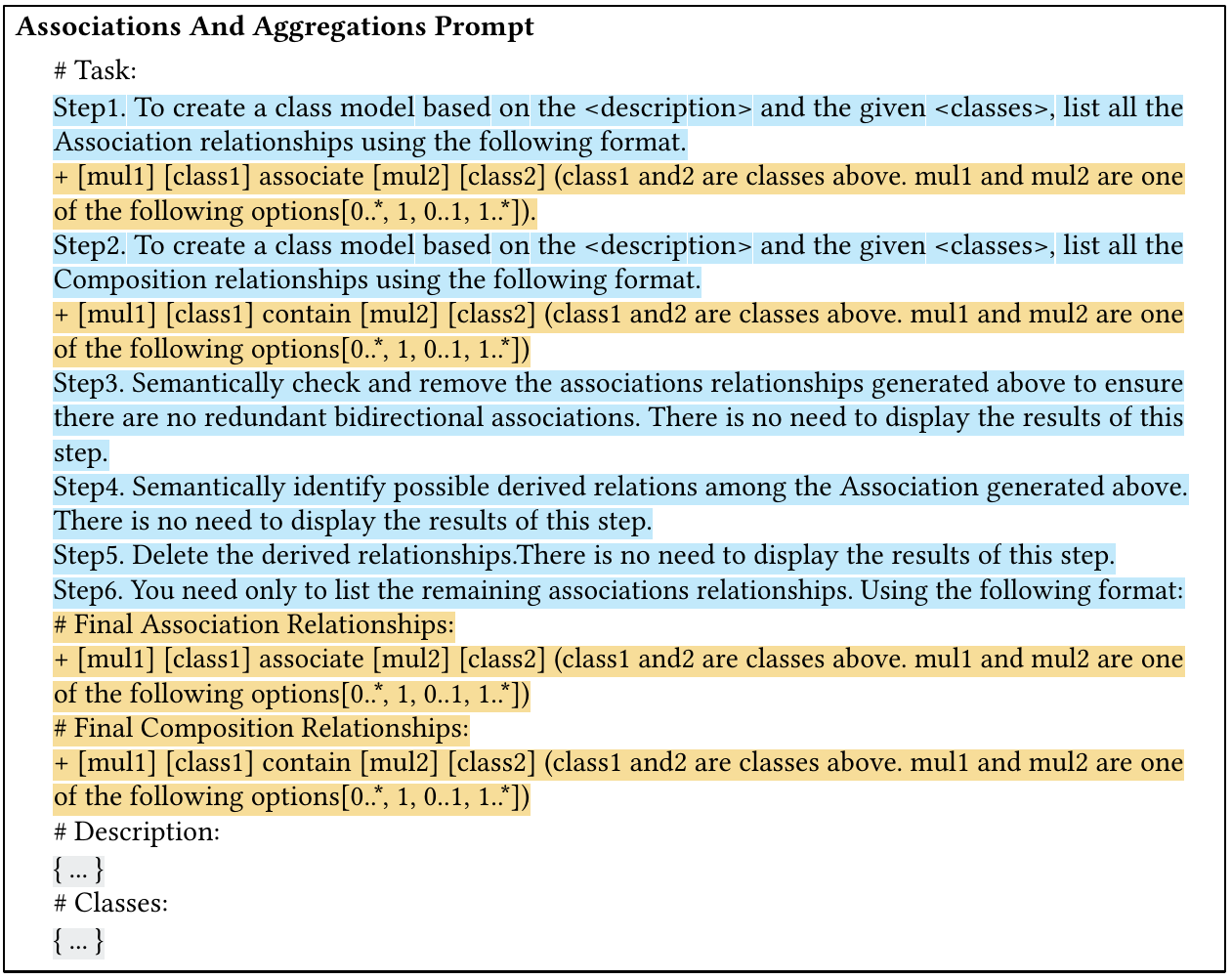}}\\
    \subfloat[The prompt for inheritances\label{fig:rel-gen-inherit}]{\includegraphics[width=\columnwidth]{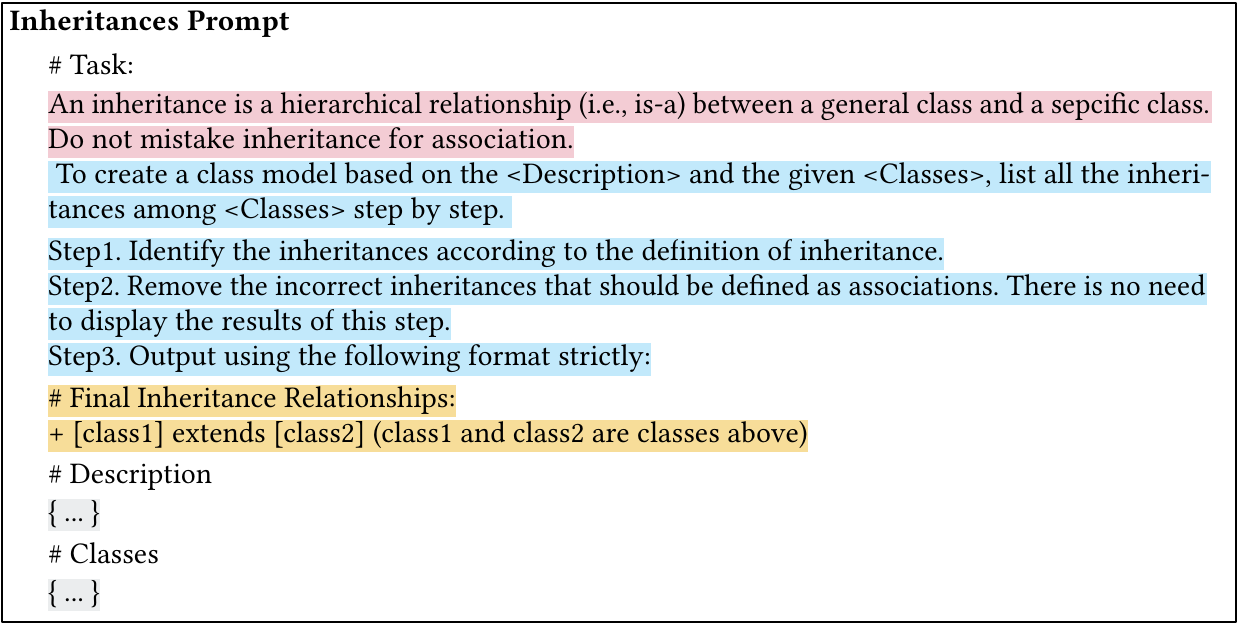}}
    \caption{Prompts of relationship generation (including \sethlcolor{taskColor}\hl{task}, \sethlcolor{formatColor}\hl{formatting}, \sethlcolor{guideColor}\hl{modeling knowledge}, and \sethlcolor{descriptionColor}\hl{system description})}
\end{figure}

Rather than addressing all types of relationships simultaneously, we again decompose the task into two independent sub-tasks, i.e., the generation of associations and aggregations, and the generation of inheritances.
Because inheritances and associations/aggregations have distinct semantics and usage in domain modeling, LLMs can devote more attention to generating particular relationships, by dividing the task into smaller ones.

\paragraph{\textbf{Association and aggregation}}
The prompt for generating associations and aggregations is shown in Figure \ref{fig:rel-gen-assoc}.
It requests LLMs to discover associations and aggregations among the given classes (obtained from class generation) based on the system description.
The prompt adopts the idea of the chain-of-thought reasoning by listing several thinking steps.
It also specifies the output formats of associations and aggregations.

\begin{remark}
Associations and aggregations are grouped into the same generation task because, e.g., in UML\footnote{The Unified Modeling Language. \url{https://www.omg.org/spec/UML/2.5/PDF}} and Ecore\footnote{Eclipse Modeling Project. \url{https://eclipse.dev/modeling/emf/}}, they are syntactically and semantically close to each other.
Based on observation, if we generate them by two separate tasks, then LLMs tend to return imprecise relationships. 
It is because an LLM cannot reconcile the results of two generation tasks.
If there is conflict in the context, the word \textit{association} may refer to both association and aggregation.
\end{remark}

\begin{figure}[!t]
    \subfloat[Combine or split association/aggregation generation\label{fig:rel-gen-composite}]{\includegraphics[width=0.75\columnwidth]{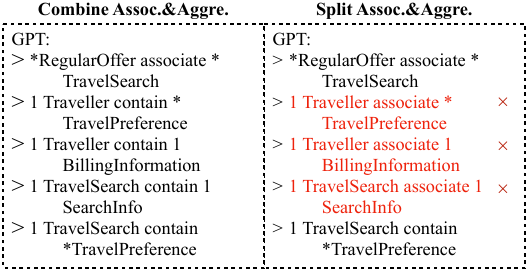}}\\
    \subfloat[Inheritance generation w/o knowledge\label{fig:rel-gen-knowledge}]{\includegraphics[width=0.75\columnwidth]{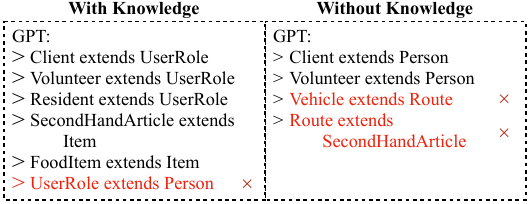}}\\
    \subfloat[Keyword selection\label{fig:rel-gen-keyword}]{\includegraphics[width=0.75\columnwidth]{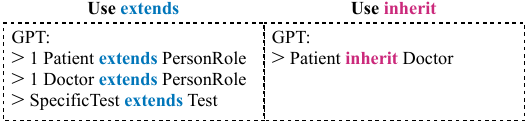}}
    \caption{Example results in pilot studies}
\end{figure}

Figure \ref{fig:rel-gen-composite} shows the result of a pilot study based on Hotel Booking Management System \cite{Chen23}.
In this system, a \emph{Traveller} owns his/her \emph{TravelPreference}, which should be viewed as an aggregation.
However, if we split the generation of association and aggregation into two separate tasks, then LLMs might mistakenly define the aggregation as an association, due to the similarity between the two concepts in certain contexts. 

\paragraph{\textbf{Inheritance}}
% The prompt for inheritances is shown in Figure \ref{fig:rel-gen-inherit}.
Figure \ref{fig:rel-gen-inherit} shows the prompt of inheritance.
It guides LLMs to identify parent-child relationships among the given classes according to the system description.
The prompt consists of three parts:
(1) the task statement;
(2) some explanations and guidelines for defining inheritances;
(3) the output format.
% other details, I need to see the prompt

\begin{remark}
    As shown in Figures \ref{fig:class-gen} and \ref{fig:rel-gen-inherit}, we inject some modeling knowledge into the prompts.
    For class generation, we tell LLMs the general rules about how to identify objects and classes.
    For inheritance generation, we tell LLMs the semantics of inheritances and associations, which are excerpted from the UML specification.
    We force LLMs to be aware of their distinctions to correctly identify inheritances.
    By providing these knowledge explicitly, we expect that LLMs improve their generation and reasoning by conducting in-context learning.
\end{remark}

Figure \ref{fig:rel-gen-knowledge} shows a typical example of our pilot study about the impact of modeling knowledge based on the Helping Hand Store (H2S) system \cite{Chen23}.
The lines in red are incorrect answers according to the referenced model while the rest are correct.
It is obvious that with the modeling knowledge provided, LLMs better understand the meaning of inheritances among classes and generate more inheritances.
In contrast, without modeling knowledge, LLMs tend to generate fewer inheritances but are more likely to confuse inheritance and association.
For instance, \textit{Vehicle} should not inherit \textit{Route} but associate with it in H2S system.

\begin{remark}
    It is worthwhile emphasizing that we carefully pick the keywords in the prompt.
    For example, in the output format, we choose \texttt{extends}, rather than \texttt{inherit} which was previously used in \cite{Chen23}.
    As we know, \texttt{extends} is a keyword representing parent-child relationships in many programming languages, such as Java, Scala, and Typescript. Since LLMs have learned from a huge amount of source code corpus during the training phase, compared with \texttt{inherit}, \texttt{extends} can stimulate LLMs to produce better answers.
    % TODO: add mini experiment
\end{remark}

Figure \ref{fig:rel-gen-keyword} shows a concrete example by switching \texttt{extends} with \texttt{inherit} when generating a Lab Requisition Management System \cite{Chen23} model.
All the three inheritances generated with the keyword \texttt{extends} match the referenced model, while the inheritance generated with the keyword \texttt{inherit} is trivially incorrect. 
Based on observation, the keyword \texttt{extends} guides LLMs to generate more reasonable inheritances, while the keyword \texttt{inherit} sometimes results in some semantic errors.

% \begin{tcolorbox}%[title = {Pilot Study}]
%     Figure \ref{} presents some examples of a pilot study that compares prompts with and without modeling knowledge.
%     It is evident that excluding modeling knowledge from the prompt results in more incorrect answers, thereby highlighting the significance of the knowledge in assisting LLMs to accurately define inheritances.
% \end{tcolorbox}

\subsection{Model Creation}\label{sec:modelcreation}
Different from the pioneering work \cite{camera}, our approach does not expect LLMs to support end-to-end model generation, even though it could be the future direction.
Instead, we ask LLMs to generate formatted lines of text during class and relationship generation.
Each formatted line provides the definition of a class/enumeration or a relationship in the model.
By parsing these formatted lines, we obtain the necessary information to create a model.

The last step of our approach, i.e. Step 4 in Figure \ref{fig:overview}, is responsible for programmatically creating a model based on the results of Step 1 and Step 3.
It is achieved as follows.
\begin{itemize}
    \item \textbf{Parsing}\quad Our approach parses the outputs of the previous generation steps according to the formats defined in the prompts. 
    However, we notice that LLMs may not strictly follow the formatting instructions and print results in certain variant formats, especially when a complex prompt is given.
    Accordingly, we construct several variant parsers to handle this issue. In case an output cannot be parsed, we ask LLMs to re-generate a new output.
    \item \textbf{Validation and fixing}\quad Even though an output of LLMs can be parsed, its content may be buggy or inconsistent. 
    Our approach must further validate the output and fix the bugs discovered.
    Presently, we employ a rule-based manner for validation and fixing by taking human knowledge into account, which checks the following rules.
    \begin{enumerate}
        \item \textit{Naming Convention}\quad A class/attribute name must conform to the \textit{camel case naming convention}. For association/aggregation names, we will create a name by joining the source and target class names.
        \item \textit{Type correctness}\quad The type of an attribute must be a data type, rather than a class. If the type of an attribute is missing, we will use \texttt{String} as the default type.
        \item \textit{Association ends}\quad The source and target of an association/aggregation relationship must be classes.
        If the relationship has a non-class end, we will try to turn it into an attribute or drop it if we cannot fix the issue.
        \item \textit{Multiplicities}\quad We check whether the multiplicity of an association/aggregation is correctly specified. 
        It not, we will use \texttt{1} as the default multiplicity.
        \item \textit{Inheritance ends}\quad LLMs may generate some parent classes that are not recognized during class generation.
        If a new parent class is required by multiple children, then we will append this class to the model.
        Otherwise, we neglect the inheritance.
    \end{enumerate}
    We do not argue that these fixing rules are sound and complete.
    Existing constraint fixing approaches may be adapted for this issue.
    We also believe that it is possible and interesting future work to investigate LLMs to fix and refine generated models, as demonstrated by \cite{Apvrille2024SystemAA}
    \item \textbf{Creation}\quad After the data is parsed and validated, we finally programmatically create an object model.
\end{itemize}

\subsection{Prototype Implementation}\label{sec:tool}
We have implemented a proof-of-concept prototype tool as an Eclipse plug-in.
The tool has a frontend and a backend.
The frontend is built upon EMF and the standard tree-like editor of Ecore.
It provides users with a GUI integration into the Ecore editor. The backend is built upon the APIs offered by OpenAI and Pyecore.
It realizes our approach proposed in this section.
The modeler can use our tool to create a model structure in a tree view based on the results of LLMs, review and choose what classes, enumerations, attributes, and relationships he/she wants to accept.
% Our tool is available online\footnote{\url{https://youtu.be/TkWQ85SmaFY}} and can be provided upon request.
% Our tool is available online \href{https://youtu.be/TkWQ85SmaFY}{here} and can be provided upon request.

% Figure \ref{fig:tool} shows screenshots of our tool.
% A modeler right-clicks an \textit{EPackage} in the editor and selects \textit{Ask LLM} to start the model generation wizard.
% Afterwards, the modeler should input/revise the system description and provide his/her access key to LLMs.
% These parameters are stored within the \textit{EPackage} as an \textit{EAnnotation}.
% If the modeler clicks \textit{Next} button, then our tool will connect to an LLM (currently, GPT 3.5-turbo) to generate a object model according to the workflow described in Figure \ref{fig:overview}.
% When the LLM returns a model, our tool lists the model structure in a tree view, where the modeler can review and choose what classes, enumerations, attributes, and relationships he/she wants to accept.
% Finally, our tool will add the modeler's selection to the selected \textit{EPackage}. 

%% file: section/evaluation.tex
\section{Evaluation}\label{sec:evaluation}
To evaluate our approach, we conduct four experiments by answering the following research questions (RQs).
\begin{itemize}
    \item \textbf{RQ1}: Does our approach improve generated models in comparison with single-prompt-based solutions?
    
    \textit{\textbf{Rationale.} The key to our approach is that domain modeling should be decomposed and addressed by several sub-questions. Hence, we must check whether our question-decomposition-based solution outperforms existing single-prompt-based solutions (see Section \ref{sec:overall-expr}).}
    
    \item \textbf{RQ2}: Does two-turn conversation outperform single-turn conversation in class generation?

    \textit{\textbf{Rationale.} To generate classes and attributes, our approach conducts a two-turn conversion with LLMs, rather than directly asking LLMs to return a list of classes from a system description. Hence, we must evaluate whether this strategy is better than direct generation (see Section \ref{sec:class-gen-expr}).}
    
    \item \textbf{RQ3}: Does the decomposition of relationship generation into two independent tasks enhance the results in comparison to a single task?

    \textit{\textbf{Rationale.} Our approach subdivides the generation of inheritance and association/aggregation into two sub-tasks. We should check whether this question decomposition improves the relationship generation
    (see Section \ref{sec:rel-gen-expr}).}

    \item \textbf{RQ4}: What temperatures should be used in our approach?

     \textit{\textbf{Rationale.} Different sub-tasks require different capabilities of LLMs. Thus, we should decide what temperature settings should be adopted for each sub-task to fully utilize the power of LLMs
    (see Section \ref{sec:temperature-expr}).}
\end{itemize}

\newcommand{\gptA}{GPT-3.5-Turbo}
\newcommand{\gptB}{GPT-4o} %% MAKE SURE
\newcommand{\llama}{Llama-3-8B} %% MAKE SURE

\paragraph{\textbf{Problem sets}} We adopt two datasets in our evaluation.
The first dataset, \textbf{\textit{DomSys}}, was provided by Chen et al. \cite{Chen23}, which includes 10 systems.
The second dataset, \textbf{\textit{NLPSet}}, is constructed by randomly selecting 10 systems provided by Bozyigit \cite{Bozyigit2024GeneratingDM}.
For each system, a textual system description and a \textit{referenced domain model} are provided.
We use the referenced models as the test Oracle to assess the qualities of generated models.
For \textit{DomSys}, Chen et al. \cite{Chen23} already defined referenced models;
for \textit{NLPSet}, since Bozyigit \cite{Bozyigit2024GeneratingDM} did not define referenced models, we ask two modeling experts to manually construct a referenced model for each system.
The basic information is listed in Table \ref{tab:problems}.

\begin{table}[!b]
    \caption{Basic information of the problem sets. S/W denotes the number of sentences/words; C/A/R denotes the number of classes/attributes/relationships.}\label{tab:problems}
    {\scriptsize{
    \begin{tabular}{p{0.2cm}|p{0.5cm}|p{3.4cm}|p{1.2cm}|p{1.3cm}}
        \hline
        \textbf{Set} & \textbf{ID} & \textbf{System name} & \textbf{Description size} & \textbf{Oracle model size}\\
       \hline
        \multirow{10}{*}{\rotatebox{-90}{\textbf{DomSys}}} &
        BTMS & Bus Transportation Management System & 18 S, 422 W & 7 C, 11 A, 9 R \\

        & H2S & The Helping Hand Store  & 22 S, 533 W & 13 C, 18 A, 18 R \\

        & LRMS & Lab Requisition Management System & 30 S, 640 W & 16 C, 43 A, 22 R \\

        & CeOS & Celebrations Organization System & 23 S, 535 W & 13 C, 23 A, 22 R \\

        & TSS & Team Sports Scouting  & 11 S, 270 W& 16 C, 24 A, 20 R \\

        & SHAS & Smart Home Automation System & 19 S, 345 W & 23 C, 26 A, 27 R \\

        & OTS & Online Tutoring System & 18 S, 341 W & 16 C, 25 A, 19 R \\

        & DBA & The Destroy Block Application & 37 S, 733 W & 15 C, 30 A, 24 R \\

        & TiOA & The Tile-O Application & 17 S, 451 W & 18 C, 19 A, 21 R \\

        & HBMS & Hotel Booking Management System & 25 S, 527 W& 18 C, 32 A, 22 R \\
        \hline
        
         \multirow{10}{*}{\rotatebox{-90}{\textbf{NLPSet}}} &
         RMS & Restaurant Management System & 20 S, 300 W & 14 C, 21 A, 11 R \\
         
         & SS & The Supermarket System & 22 S, 657 W & 5 C, 15 A, 6 R \\
         
         & HRS & HotelReservation System & 20 S, 362 W & 4 C, 10 A, 2 R \\
         
         & RIRS & Romano's Italian Restaurant System & 60 S, 1002 W & 17 C, 51 A, 15 R \\ 
         
         &  VRS & SchlockBuster Video Rental Store & 48 S, 668 W & 13 C, 35 A, 17 R \\
         
         & BoG & Board Game & 25 S, 287 W & 6 C, 10 A, 6 R \\
         
         & CS & Centralized Store & 24 S, 431 W & 8 C, 8 A, 6 R \\
         
         & SLS & The Stuff Lending System & 28 S, 475 W & 8 C, 15 A, 11 R \\
         
         & HRIS & Human-Resources Information System & 26 S, 539 W & 9 C, 26 A, 4 R \\
         
         & YCS & The Yacht Club System & 31 S, 619 W & 12 C, 21 A, 14 R \\
         
         \hline
    \end{tabular}
    }}
\end{table}
% The baseline experiment uses a problem description as input to LLMs, utilizes a single conversation to generate classes and relationships. Subsequently, we assess the quality of the generated classes and relationships by comparing them with the reference solution. To ensure the experiment's validity, we conducted fifty times experiments for each problem.

% There are some differences between our approach and baseline. Our approach is to decompose complex questions, change prompt strategy of sub-question and merge all sub-solution. To ensure the method's validity and the experiment's fairness, parameter settings of LLMs are same with baseline. 

\paragraph{\textbf{Metrics}} To estimate how close a generated model and the corresponding Oracle model are, we follow the basic idea of \cite{Chen23} that measures the similarity of classes, attributes, and relationships in terms of precision ($Prec$), recall ($Rec$), and F1 score ($F1$).

To calculate the metrics, we use the following equations:
\begin{equation}
    Prec=\frac{TP}{TP+FP}
\end{equation}
\begin{equation}
    Rec=\frac{TP}{TP+FN}
\end{equation}
\begin{equation}
    F1=\frac{2\times Prec\times Rec}{Prec+Rec}
\end{equation}
where $TP$, $FP$, and $FN$ refer to the numbers of true positives (i.e., generated answers that match Oracle), false positives (i.e., generated answers that do not match Oracle), and false negatives (i.e., Oracle answers that are not generated), respectively.

\paragraph{\textbf{Model matching}}
To evaluate the quality of generated models using the above metrics, 
we manually compare the generated models with the Oracle models. 
We mainly adopt the matching strategy described in Chen's paper\cite{Chen23}. 
Nevertheless, some of their matching rules are subjective, requiring in-depth semantic matching. 
To eliminate any bias, we make some modifications to the matching rules, focusing on more objective criteria. 
The adapted matching rules can be summarized as follows.

% We evaluate the quality of generated models by calculating the metrics mentioned above. 
% Moreover, we manually compare the generated models with the Oracle models.
% % We determine whether a generated model can \textit{match} on each task.
% % We match the generated model and the Oracle model by manual inspection.
% % Our method is consistent with that of Chen et al. \cite{Chen23} to match models.
% In this regard, we adopt Chen's matching approach, and the two approaches are consistent.
% However, there are some rules and subjective matching in Chen's approach.
% To ensure the fairness of the experiment and reduce subjective bias, we make some adaptions, ignoring subjective matching in our approach, and using strict matching rules summarized as follows: 
\begin{itemize}
 % For two classes, if their names are similar (with a 1-gram similarity higher than 0.9), then they are paired.
    \item For two classes, if their names are similar (with a 1-gram similarity higher than 0.9), then they are paired. 
    Additionally, if class names are partially the same and refer to the same concept (e.g., Bus vs. BusVehicle, Schedule vs. DriverSchedule), then they are paired.
    \item For two attributes, if their owner classes have been matched, their names are similar, then they are paired.
    \item For two inheritances, if their parent classes and child classes have matched, respectively, and their relationship type is compatible, then they are paired.
    \item For two associations/aggregations (\textit{s1} - \textit{t1}, \textit{s2} - \textit{t2}), in general cases, if \textit{s1} matches \textit{s2} and \textit{t1} matches  \textit{t2}, then they are paired. 
    % However, in cases where bidirectional relationships exist(e.g. 1 Ingredient associate 1 FoodItem), if the generated model reflects the inverse relationship(e.g. 1 FoodItem associate 1 Ingredient), that they are also matched.
    However, because associations can be bidirectional, if \textit{s1} matches  \textit{t2} and \textit{s2} matches  \textit{t1}, they are also matched. 
    For example, ``1 Ingredient associates 1 FoodItem'' is correct.
    If the generated is the reverse relationship (i.e. `1 FoodItem associates 1 Ingredient'), then they are also matched.
    % \item Manual inspection will further be conducted to find semantic matches.
    % % % add part:
    % % Class names partially match elements of the same type in the reference model.
    % % It is important to note that this manual match is based on string comparison rather than semantics.
    % % % 
    % % For example, two classes may own different names (e.g. Bus vs. BusVehicle, Schedule vs. DriverSchedule) but are a partial match and refer to the same concept; two associations may be viewed as the two directions of the same bidirectional relationships.
    % For each generated model, at least two of the authors will check the matches together to coordinate disagreements.
    % Different from \cite{Chen23}, we ignore \textit{partial matches} to avoid subjective bias.
\end{itemize}

% Due to the large number of tests, the \textit{manual inspection} (R4) is unaffordable in our evaluation. 
% To avoid human intervention, we employ an automated inspection (say R4') for unmatched classes to take the place of R4. 
% If two classes $c_1$ and $c_2$ cannot be matched by R1, then we further calculate their Jaccard similarity (see Equation (\ref{eq:jaccard})) by comparing their attributes, as is inspired by exiting model matching approaches \cite{nwm,snwm}.
% We use R2 to determine whether two attributes are equal, while ignoring their owner classes.
% When $Jaccard(c_1,c_2)$ is higher than a threshold, we regard that $c_1$ and $c_2$ are semantically matched.
% In practice, the threshold value is empirically set to 0.6.
% \begin{equation}
%     Jaccard(c_1,c_2) =  \frac{| c_1\text{'s attributes} \cap c_2\text{'s attributes}|}{|c_1\text{'s attributes} \cup c_2\text{'s attributes}|}
%     \label{eq:jaccard}
% \end{equation}

% \textit{\textbf{Implementation details.}} 
% 1).使用的模型，模型参数的设置。
% 2).强调在所有的评估对比实验中，我们跳过了 'integration and Model Creation' 步骤。 => 我觉得还是放在threats to validity中讨论，我们可以统计一下我们的方法中有多少用到了默认值，我认为现在很少用到，可以忽略
% 3).说明为什么不用现有模型比较方法。

% \subsection{RQ1: Does our approach generate better concept models compared with the baseline?}
\subsection{Experiment 1: overall effectiveness}\label{sec:overall-expr}
The goal of the 1st experiment is to evaluate the effectiveness of our LLM-based approach to domain modeling.
In this experiment, we compare our approach with Chen et al. \cite{Chen23}, which is regarded as a representative single-prompt-based domain model generation approach.
% For a fair comparison, we only try their 0-shot and \textcolor{red}{CoT} versions because our approach uses 0-shot \textcolor{red}{and CoT} prompts. % should we try their CoT version?
For a fair comparison, we adopt Chen's 0-shot version, as our approach utilizes both 0-shot and Chain of Thought (CoT) prompts. However, it's important to note that Chen's CoT prompt does not include step guidelines, which differs from ours and cannot be fairly compared.

\begin{table*}[!tb]
    \caption{Average precision, recall, and F1 scores of Experiment 1}\label{tab:1st-exp-sys}
    {\small{
    \begin{tabular}{c||c||c|cccc|cccc|cccc}
        \hline
        \multirow{2}{*}{\textbf{LLM}} & \multirow{2}{*}{\textbf{Dataset}} & \multirow{2}{*}{\textbf{Approach}} & \multicolumn{4}{c|}{\textbf{Precision}} & \multicolumn{4}{c|}{\textbf{Recall}} & \multicolumn{4}{c}{\textbf{F1 score}} \\
        & & & Class & Attr. & Inher. & Assoc. & Class & Attr. & Inher. & Assoc. & Class & Attr. & Inher. & Assoc.\\
        \hline

        \multirow{6}{*}{\rotatebox{0}{\gptA}} & \multirow{2}{*}{DomSys} & Ours & 0.808 & 0.222 & \textbf{0.001} & \textbf{0.327} & \textbf{0.501} & \textbf{0.259} & \textbf{0.003} & \textbf{0.142} & \textbf{0.608} & \textbf{0.234} & \textbf{0.003} & \textbf{0.191} \\
        & & Chen-0-Shot & \textbf{0.825} & \textbf{0.246} & \textbf{0.001} & 0.196 & 0.419 & 0.179 & 0.001 & 0.072 & 0.542 & 0.200 & 0.001 & 0.099 \\
        % & & Chen-CoT & 0.000 & 0.000 & 0.000 & 0.000 & 0.000 & 0.000 & 0.000 & 0.000 & 0.000 & 0.000 & 0.000 & 0.000 \\
        \cline{2-15}
        & \multirow{3}{*}{NLPSet} & Ours & 0.654 & 0.312 & 0.103 & \textbf{0.308} & 0.612 & \textbf{0.384} & 0.114 & \textbf{0.308} & 0.611 & \textbf{0.338} & 0.104 & \textbf{0.246} \\
        & & Chen-0-Shot & 0.575 & \textbf{0.349} & 0.014 & 0.192 & 0.518 & 0.272 & 0.01 & 0.134 & 0.525 & 0.287 & 0.011 & 0.138  \\
        % & & Chen-CoT & 0.000 & 0.000 & 0.000 & 0.000 & 0.000 & 0.000 & 0.000 & 0.000 & 0.000 & 0.000 & 0.000 & 0.000 \\
        \hline

        % \multirow{6}{*}{\rotatebox{0}{\gptB}} & \multirow{2}{*}{DomSys} & Ours & 0.766 & 0.210 & 0.217 & 0.020 & 0.562 & 0.332 & 0.186 & 0.003 & 0.641 & 0.253 & 0.193 & 0.004 \\
        % & & Chen-0-Shot & 0.776 & 0.257 & 0.223 & 0.011 & 0.503 & 0.260 & 0.119 & 0.002 & 0.603 & 0.255 & 0.150 & 0.004 \\
        % % & & Chen-CoT & 0.000 & 0.000 & 0.000 & 0.000 & 0.000 & 0.000 & 0.000 & 0.000 & 0.000 & 0.000 & 0.000 & 0.000 \\
        % \cline{2-15}
        % & \multirow{3}{*}{NLPSet} & Ours & 0.602 & 0.243 & 0.188 & 0.060 & 0.674 & 0.432 & 0.323 & 0.076 & 0.619 & 0.302 & 0.210 & 0.065 \\
        % & & Chen-0-Shot & 0.557 & 0.271 & 0.272 & 0.032 & 0.643 & 0.339 & 0.319 & 0.071 & 0.580 & 0.294 & 0.266 & 0.037 \\
        % % & & Chen-CoT & 0.000 & 0.000 & 0.000 & 0.000 & 0.000 & 0.000 & 0.000 & 0.000 & 0.000 & 0.000 & 0.000 & 0.000 \\
        % \hline

        \multirow{6}{*}{\rotatebox{0}{\llama}} & \multirow{2}{*}{DomSys} & Ours & 0.772 & 0.183 & 0.000 & 0.229 & 0.431 & 0.229 & 0.000 & 0.084 & 0.546 & 0.197 & 0.000 &  0.122 \\
        & & Chen-0-Shot & 0.718 & 0.146 & 0.000 & 0.234 & 0.451 & 0.164 & 0.000 & 0.072 & 0.550 & 0.151 & 0.000 & 0.108 \\
        % & & Chen-CoT & 0.000 & 0.000 & 0.000 & 0.000 & 0.000 & 0.000 & 0.000 & 0.000 & 0.000 & 0.000 & 0.000 & 0.000 \\
        \cline{2-15}
        & \multirow{3}{*}{NLPSet} & Ours & \textbf{0.656} & 0.232 & \textbf{0.233} & 0.198 & \textbf{0.630} & 0.349 & \textbf{0.183} & 0.193 & \textbf{0.625} & 0.272 & \textbf{0.190} & 0.181  \\
        & & Chen-0-Shot & 0.626 & 0.184 & 0.150 & 0.184 & 0.529 & 0.141 & 0.150 & 0.163 & 0.554 & 0.128 & 0.150 & 0.165 \\
        % & & Chen-CoT & 0.000 & 0.000 & 0.000 & 0.000 & 0.000 & 0.000 & 0.000 & 0.000 & 0.000 & 0.000 & 0.000 & 0.000 \\
        \hline
     
    \end{tabular}
    }}
    \begin{flushright}
        \footnotesize{The bold data indicate the best scores for each task under different models for the same dataset.}
    \end{flushright}
\end{table*}

\begin{figure*}
\overviewbarplot{Class@\gptA}
{BTMS,TSS,SHAS,TiOA,DBA,OTS,HBMS,H2S,CeOS,LRMS,RMS,SS,HRS,RIRS,VRS,BoG,CS,SLS,HRIS,YCS}
{(1,0.791)(2,0.7)(3,0.499)(4,0.484)(5,0.583)(6,0.551)(7,0.573)(8,0.541)(9,0.709)(10,0.65)(11,0.571)(12,0.203)(13,0.765)(14,0.683)(15,0.827)(16,0.728)(17,0.601)(18,0.572)(19,0.694)(20,0.467)}
{(1,0.775)(2,0.54)(3,0.513)(4,0.437)(5,0.555)(6,0.563)(7,0.44)(8,0.503)(9,0.606)(10,0.485)(11,0.488)(12,0.195)(13,0.642)(14,0.624)(15,0.795)(16,0.443)(17,0.463)(18,0.472)(19,0.644)(20,0.484)}
\overviewbarplot{Attribute@\gptA}{BTMS,TSS,SHAS,TiOA,DBA,OTS,HBMS,H2S,CeOS,LRMS,RMS,SS,HRS,RIRS,VRS,BoG,CS,SLS,HRIS,YCS}
{(1,0.497)(2,0.24)(3,0.138)(4,0.251)(5,0.167)(6,0.073)(7,0.348)(8,0.132)(9,0.24)(10,0.258)(11,0.347)(12,0.155)(13,0.248)(14,0.639)(15,0.518)(16,0.305)(17,0.135)(18,0.288)(19,0.357)(20,0.385)}
{(1,0.431)(2,0.182)(3,0.261)(4,0.011)(5,0.18)(6,0.156)(7,0.237)(8,0.1)(9,0.188)(10,0.249)(11,0.353)(12,0.093)(13,0.096)(14,0.538)(15,0.409)(16,0.451)(17,0.122)(18,0.167)(19,0.233)(20,0.405)}
\overviewbarplot{Inheritance@\gptA}{BTMS*,TSS,SHAS*,TiOA,DBA*,OTS,HBMS,H2S,CeOS,LRMS,RMS*,SS*,HRS*,RIRS,VRS*,BoG*,CS,SLS,HRIS,YCS}
{(1,0.0)(2,0.0)(3,0.0)(4,0.026)(5,0.0)(6,0.0)(7,0.0)(8,0.006)(9,0.0)(10,0.0)(11,0.0)(12,0.0)(13,0.0)(14,0.0)(15,0.0)(16,0.0)(17,0.107)(18,0.79)(19,0.1)(20,0.04)}
{(1,0.0)(2,0.0)(3,0.0)(4,0.007)(5,0.0)(6,0.0)(7,0.0)(8,0.0)(9,0.0)(10,0.0)(11,0.0)(12,0.0)(13,0.0)(14,0.0)(15,0.0)(16,0.0)(17,0.033)(18,0.047)(19,0.033)(20,0.0)}
\overviewbarplot{Association@\gptA}{BTMS,TSS,SHAS,TiOA,DBA,OTS,HBMS,H2S,CeOS,LRMS,RMS,SS,HRS,RIRS,VRS,BoG,CS,SLS,HRIS,YCS}
{(1,0.14)(2,0.114)(3,0.307)(4,0.228)(5,0.352)(6,0.062)(7,0.113)(8,0.128)(9,0.237)(10,0.224)(11,0.063)(12,0.0)(13,0.688)(14,0.186)(15,0.18)(16,0.371)(17,0.193)(18,0.193)(19,0.495)(20,0.092)}
{(1,0.077)(2,0.043)(3,0.227)(4,0.122)(5,0.217)(6,0.123)(7,0.023)(8,0.016)(9,0.097)(10,0.046)(11,0.09)(12,0.0)(13,0.316)(14,0.103)(15,0.299)(16,0.229)(17,0.07)(18,0.094)(19,0.092)(20,0.087)}

\overviewbarplot{Class@\llama}{BTMS,TSS,SHAS,TiOA,DBA,OTS,HBMS,H2S,CeOS,LRMS,RMS,SS,HRS,RIRS,VRS,BoG,CS,SLS,HRIS,YCS}
{(1,0.615)(2,0.72)(3,0.516)(4,0.444)(5,0.4)(6,0.667)(7,0.37)(8,0.476)(9,0.667)(10,0.583)(11,0.615)(12,0.267)(13,0.727)(14,0.593)(15,0.8)(16,0.727)(17,0.625)(18,0.533)(19,0.75)(20,0.615)}
{(1,0.615)(2,0.64)(3,0.649)(4,0.444)(5,0.455)(6,0.643)(7,0.452)(8,0.435)(9,0.5)(10,0.667)(11,0.519)(12,0.235)(13,0.8)(14,0.333)(15,0.727)(16,0.6)(17,0.462)(18,0.615)(19,0.667)(20,0.583)}
\overviewbarplot{Attribute@\llama}{BTMS,TSS,SHAS,TiOA,DBA,OTS,HBMS,H2S,CeOS,LRMS,RMS,SS,HRS,RIRS,VRS,BoG,CS,SLS,HRIS,YCS}
{(1,0.5)(2,0.292)(3,0.073)(4,0.048)(5,0.222)(6,0.203)(7,0.059)(8,0.197)(9,0.074)(10,0.304)(11,0.364)(12,0.103)(13,0.0)(14,0.298)(15,0.564)(16,0.552)(17,0.118)(18,0.089)(19,0.25)(20,0.386)}
{(1,0.435)(2,0.051)(3,0.033)(4,0.213)(5,0.105)(6,0.097)(7,0.069)(8,0.158)(9,0.115)(10,0.234)(11,0.217)(12,0.0)(13,0.194)(14,0.175)(15,0.246)(16,0.0)(17,0.0)(18,0.0)(19,0.128)(20,0.316)}
\overviewbarplot{Inheritance@\llama}{BTMS*,TSS,SHAS*,TiOA,DBA*,OTS,HBMS,H2S,CeOS,LRMS,RMS*,SS*,HRS*,RIRS,VRS*,BoG*,CS,SLS,HRIS,YCS}
{(1,0.0)(2,0.0)(3,0.0)(4,0.0)(5,0.0)(6,0.0)(7,0.0)(8,0.0)(9,0.0)(10,0.0)(11,0.0)(12,0.0)(13,0.0)(14,0.0)(15,0.0)(16,0.0)(17,0.4)(18,0.0)(19,1.0)(20,0.5)}
{(1,0.0)(2,0.0)(3,0.0)(4,0.0)(5,0.0)(6,0.0)(7,0.0)(8,0.0)(9,0.0)(10,0.0)(11,0.0)(12,0.0)(13,0.0)(14,0.0)(15,0.0)(16,0.0)(17,0.5)(18,0.0)(19,1.0)(20,0.0)}
\overviewbarplot{Association@\llama}{BTMS,TSS,SHAS,TiOA,DBA,OTS,HBMS,H2S,CeOS,LRMS,RMS,SS,HRS,RIRS,VRS,BoG,CS,SLS,HRIS,YCS}
{(1,0.0)(2,0.083)(3,0.188)(4,0.133)(5,0.174)(6,0.25)(7,0.0)(8,0.0)(9,0.24)(10,0.148)(11,0.087)(12,0.0)(13,0.0)(14,0.19)(15,0.37)(16,0.364)(17,0.2)(18,0.353)(19,0.25)(20,0.0)}
{(1,0.0)(2,0.167)(3,0.071)(4,0.235)(5,0.25)(6,0.19)(7,0.0)(8,0.0)(9,0.0)(10,0.167)(11,0.25)(12,0.0)(13,0.0)(14,0.087)(15,0.0)(16,0.4)(17,0.0)(18,0.353)(19,0.333)(20,0.222)}
\vspace{-0.3cm} % 调整垂直间距
\makebox[\textwidth][r]{\footnotesize * represents that the system has no inheritance in the oracle model.} % 使用makebox使其居中显示

\caption{F1 scores of all systems}\label{fig:f1-details-exp1}
\end{figure*}

\paragraph{Process} 
We ask our approach and Chen's approach to generate object-oriented domain models for all 20 systems in the two problem sets.
For each system, both approaches are required to generate 50 models on \gptA{} and \llama{}.
We verify the generated models with Oracle models and report the precision, recall, and F1 score of each approach.
For Chen's approach, we used the original parameter settings provided by \cite{Chen23}.
For our approach, we choose temperature 0.4 for classes generation, 0.9 for associations/aggregations generation, and 0.8 for inheritances generation (see Section \ref{sec:temperature-expr} for the details about temperature selection).

\paragraph{Results}
% Figure \ref{fig:1st-exp-overall} presents the F1 scores of the first experiment for different generation tasks (i.e., class, attribute, and relationship generation).
% The X-axis denotes the generation task, and the Y-axis denotes F1 score.
% The boxes in red denote our approach, while those in black denote the baseline.
% Figure \ref{fig:1st-exp-overall} shows the following facts.
% \begin{itemize}
%     \item For classes, the F1 score of our approach varies from XXX to YYY, with a mean of 0.45, while the F1 score of the baseline varies from XXX to YYY, with a mean of 0.40.
%     \item For attributes, the F1 score of our approach varies from XXX to YYY, with a mean of 0.20, while the F1 score of the baseline varies from XXX to YYY, with a mean of 0.13.
%     \item For relationships, the F1 score of our approach varies from XXX to YYY, with a mean of 0.34, while the F1 score of the baseline varies from XXX to YYY, with a mean of ZZZ.
% \end{itemize}
% Second, I need a wide table to present the detailed results. There are & & 7 Columns. The 1st column lists the system names. The 2nd-4th columns show the f1 of our classes, attributes, and relationships; 5th-7th show f1 of baseline's.

Table \ref{tab:1st-exp-sys} shows the average precision, recall, and F1 scores of different generation tasks for each LLM, dataset, and type of modeling element for our approach and baseline approach. 
% It can be observed that our approach outperforms Chen's approach for each generation task, each LLM, on F1 and recall score.
It can be observed that our approach outperforms the baseline in terms of F1 score on the task, especially on \gptA{}, raising 12.18\% (class), 17\% (attribute), 200\% (inheritance) and 92.93\% (association) on DomSys, and raising 16.38\% (class), 17.77\% (attribute), 845.45\% (inheritance) and 78.26\% (association) on NLPSet.

% First, let us focus on F1 score. 
% For class generation, our approach outperforms the baseline in 9 out of 10 systems, and the only exception is DBA.
% For attribute and relationship generation, our approach is better than the baseline in all the 10 systems.
% Regarding the other two metrics, our approach improves recall more than precision for classes, attributes, and relationships.
% It is because our approach usually generates more answers compared with the baseline, resulting in the increasing of recall for all tasks and the decrease of precision for class generation.

% First, let us focus on F1 score. 
% Our approach is better than the baseline in all systems for each generation task.
% \textcolor{red}{For class generation, our approach outperforms the baseline in 17 out of 20 systems, and the exceptions are SHAS, OTS and YCS.}
% For attribute and relationship generation, our approach is better than the baseline in all the 10 systems.
% Regarding the other two metrics, our approach improves recall more than precision for classes, attributes, and relationships.
% It is because our approach usually generates more answers compared with the baseline, resulting in the increasing of recall for all tasks and the decrease of precision for class generation.
% for example, ...

% Third, I need ten box plots, each of which depicts the results of a certain system. Each box plot contains six boxes representing recall, prec, f1 of ours and baselines

\tcbset{
    colback=gray!10,      % 背景色
    colframe=black,     % 边框颜色
    boxrule=1pt,      % 边框粗细
    width=\columnwidth, % 框宽度与栏目宽度一致
    arc=2mm,            % 边框圆角
    outer arc=2mm,      % 外圆角
    boxsep=5pt,         % 框内边距
    left=5pt,           % 左边距
    right=5pt,          % 右边距
    top=5pt,            % 上边距
    bottom=5pt,         % 下边距
}

\begin{tcolorbox}
\textbf{\textit{Answer to RQ1.}} According to Table \ref{tab:1st-exp-sys}, our answer to \textbf{RQ1} is \textbf{YES}---our approach, by using the idea of question decomposition, outperforms the single-prompt-based solution. 
Since our approach tends to improve recall, we believe that it can be used as a modeling assistant to suggest classes, attributes, and relationships that might be neglected by a human modeler.
\end{tcolorbox}

% The results show that LLMs generate classes from system descriptions better than attributes, they generate attributes better than associations, and they generate associations better than inheritances.
Figure \ref{fig:f1-details-exp1} presents a comparison of the average F1 score for different generation tasks against the baseline across all systems, utilizing \gptA{} and \llama{}.
Our approach achieves more \textit{best scores} (see the bold data in Table \ref{tab:1st-exp-sys}) than Chen's approach.
In general, for both classes and attributes, our approach outperforms over.
For relationships, it does not better than classes and attributes. 
% We analyze the reasons are two-fold.
We analyze the reasons from the following two aspects.
Firstly, generating relationships from descriptions is inherently challenging, and requires the information extraction and analysis capabilities of LLMs.
Secondly, classes generation impacts relationships generation.(see discussion in section \ref{sec:rel-gen-expr})
For example, if the parent class is not correctly generated, the inheritance relationship is not matched.

We found that LLMs generate associations better than inheritances.
% This can only be partially attributed to two aspects.
The partial reasons can be attributed to the following two factors.
% Firstly, there is no inheritance in the Oracle model of some systems (e.g., BTMS, SHAS, BoG).
Firstly, some Oracle models, such as BTMS, SHAS, and BoG, do not include any inheritance.
% Hence, these systems score zero on inheritance generation, contributing to a relatively lower overall score for inheritance.
As a result, for these systems, the inheritance scores are always zero, decreasing the overall results for inheritance.
% Secondly, % insufficient system samples result in low scores.
Secondly, the total number of inheritances in Oracle models is small. 
Hence, we may not have enough samples in this experiment.

% 通过观察table2的加粗数据，关于其他的两个指标，我们的方法提高recall比precision要好一些.
% 此外，对于我们的方法，我们观察模型A在属性和关联的生成上比模型B要好一些。
% 因为属性和关联的最好分数均由gpt3.5获得。
% 但是，对于类的生成，llama3-8b的f1 比 gpt3.5高一点。
% 对于继承，llama3-8b的precision、recall, and f1都比gpt3.5高。
% 因此，我们we believe that different LLMs may be better suited for distinct tasks.

Regarding the other two metrics, our approach significantly improves recall than precision for classes, attributes, and relationships.
% Because our method also generates some incorrect answers besides correct answers.
% But these answers can be seen as modeling suggests to help modelers to model.
% For example, in the BTMS system, our approach would generate enumeration (i.e., `BusStatus(Normal, RepairShop): Indicates the status of a bus'), which is not generated in baseline. 
Furthermore, we observe that \gptA{} outperforms \llama{} for attributes and associations generation, since \gptA{} has the best scores for both generations.
However, for inheritances generation, \llama{} outperforms \gptA{}, since \llama{} gets higher scores in precision, recall, and F1 scores on NLPSet. 
Therefore, we believe that different LLMs may have different advantages in solving tasks at different levels.

% Regarding the other two metrics, our approach significantly improves recall than precision for classes, attributes, and relationships.
% Furthermore, we observe that \gptA{} outperforms \llama{} for attributes and associations generation tasks. 
% \gptA{} has the best scores for attributes and association in precision, recall, and f1 scores.
% While \llama{} outperforms \gptA{} in the inheritance generation task.
% For class, \gptA{} has best scores in DomSys, \llama{} has best scores in NLPSet.
% Therefore, we believe that different LLMs may be better suited for distinct tasks.

When doing this experiment, we observed that the output format of the baseline approach was very unstable.
It seems that \gptA{} did not follow the format specified in the prompt of the baseline approach.
The following shows some variants of the generated relationships.

\begin{small}
\begin{verbatim}
- 1 Player can be shortlisted in ShortListedPlayers
- TutoringSession may be canceled by 1 Tutor or 1 Student
- [1..*] Lab offer 0..* Test (They are associations)
- 1..* Traveller associate Hotel 0..*
\end{verbatim}
\end{small}

% To control the output format, 
% we tried a few solutions, including (1) replacing the baseline's format specification with ours \textcolor{red}{and (2) providing some examples ??}
% We managed to improve class generation, but failed to control relationship generation.
% Consequently, we manually examined the relationships generated by the baseline and strictly followed R3 to fill the last three columns of Table \ref{tab:1st-exp-sys}.
We think that the reason why the baseline cannot control its output is twofold.
Firstly, the randomness of LLMs might produce a randomized output format.
However, we ran our approach and the baseline approach alternately during this experiment, and our approach can have a stable output format.
Secondly, LLMs might not consider all the requirements when answering a complex question (e.g., generating a full object model).
Nevertheless, our approach breaks down the complex question into smaller ones following real-world modeling guidelines.
% so that LLMs can easily concentrate on a specific aspect and follow our instructions.
This strategy enables LLMs to focus intently on specific aspects and adhere closely to the instructions.
% We assume that the ability to adhere to the output format is an essential property for LLM-based approaches, which ensures that the generated output is algorithmically processable.

\subsection{Experiment 2: class generation strategy}\label{sec:class-gen-expr}
We adopt a 2-turn conversation-based approach for class generation to allow an LLM to prepare a draft first and later refine its answer.
The goal of the 2nd experiment is to evaluate the effectiveness of this strategy.
We mainly focus on the following research question:
\begin{itemize}
    \item \textbf{RQ2.} \textit{Does 2-turn conversation outperform single-turn conversation in class generation?}
\end{itemize}

\begin{figure}[!tb]
    \centering
    \includegraphics[width=0.9\columnwidth]{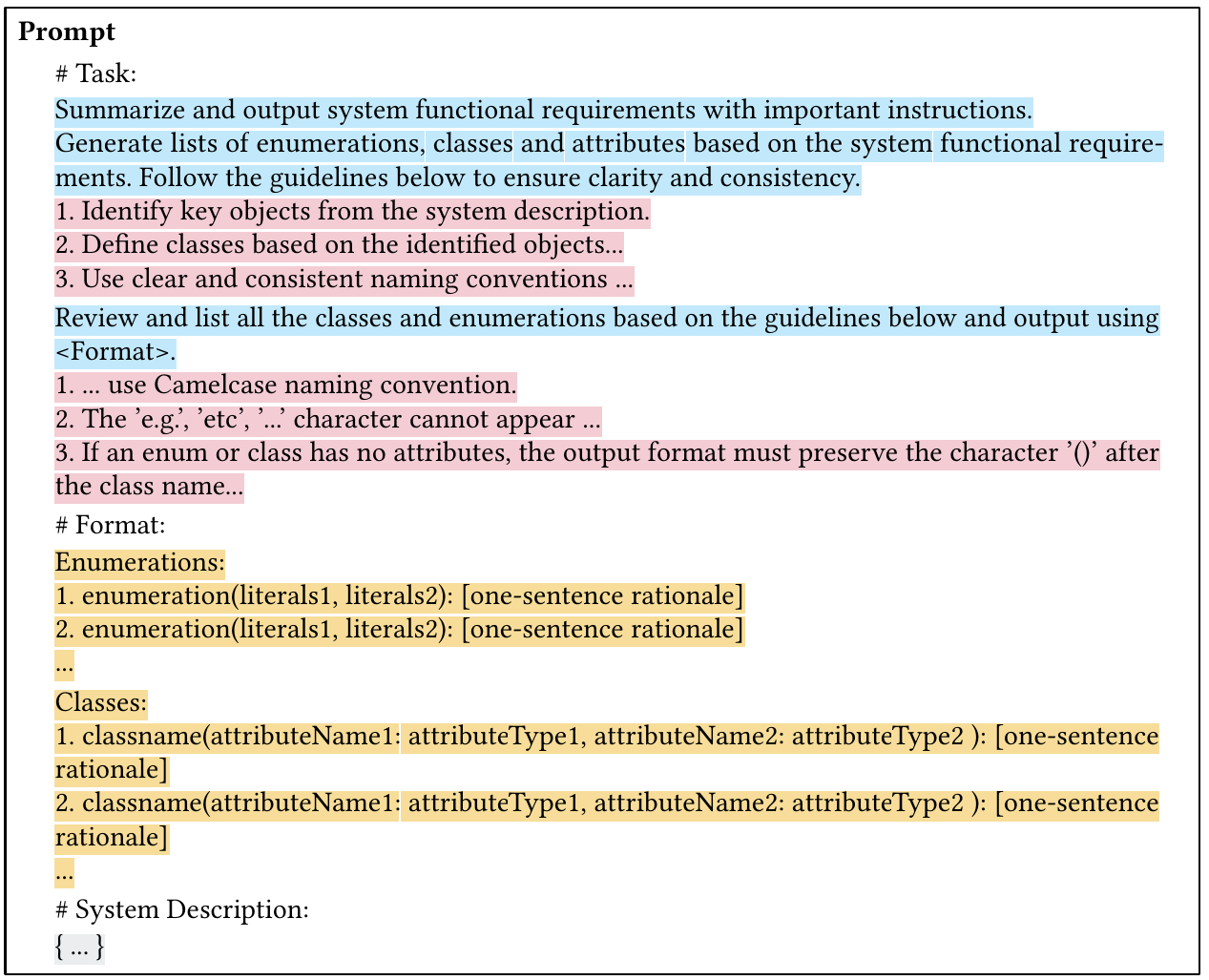}
    \caption{The prompt of single-turn conversation for classes}
    \label{fig:exp2-single-prompt}
\end{figure}

\paragraph{Process} We set up a single-turn conversation by merging the two prompts for class generation in our approach as the baseline of the 2nd experiment, as shown in Figure \ref{fig:exp2-single-prompt}.
We ask our approach and the baseline to generate classes, attributes, and enumerations for all 20 systems in the problem set.
For each system, both approaches are required to generate 50 models on \gptA{}.
By comparing the generated results with Oracle models, we calculate precision, recall, and F1 score of each approach.
In this experiment, we choose the temperature 0.4 for both approaches.

\begin{figure}[!tb]
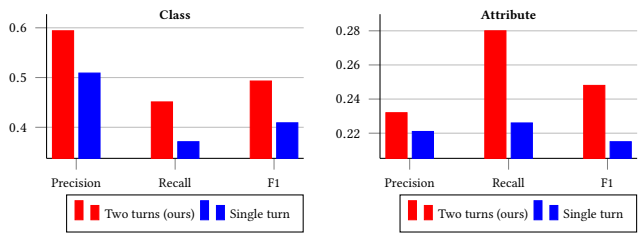

    \centering
    \compbarplot{Class}{(1, 0.594) (2, 0.451) (3, 0.493)}{(1, 0.509) (2, 0.371) (3, 0.409)}{Two turns (ours), Single turn}
    \hspace{0.5em}
    \compbarplot{Attribute}{(1, 0.232) (2, 0.280) (3, 0.248)}{(1, 0.221) (2, 0.226) (3, 0.215)}{Two turns (ours), Single turn}
    \caption{Comparison of class generation strategies}
    \label{fig:2nd-exp}
\end{figure}

\paragraph{Results}
% I need two bar plots for classes and attributes. Each bar plot contains 6 bars, denoting prec, recall, f1 of ours and baselines.
Figure \ref{fig:2nd-exp} illustrates comparison of precision, recall, and F1 score between our approach and the single-turn approach for classes and attributes.
\textbf{For classes,} the average precision, recall, and F1 score of our approach are around 0.59, 0.45, and 0.49, while those of the single-turn approach are about 0.51, 0.37, and 0.41, achieving an average improvement of 16.70\%  in precision, 21.56\%  in recall, and 20.54\%  in the F1 score. 
\textbf{For attributes,} the average precision, recall, and F1 score of our approach are around 0.23, 0.28, and 0.25, while those of the single-turn approach are about 0.22, 0.23, and 0.22, achieving an average improvement of 4.98\% in precision, 23.89\% in recall, and 15.35\% in the F1 score.
Our approach is better than the single-turn approach in terms of all the metrics.

\begin{tcolorbox}
\textbf{\textit{Answer to RQ2.}} According to Figure \ref{fig:2nd-exp}, our answer to \textbf{RQ2} is \textbf{YES}---the 2-turn conversation strategy outperforms the single-turn one in class generation.
Particularly, we find that 2-turn conversation significantly improves class generation and sightly enhances attribute generation.
The 2-turn conversation strategy is effective, especially for classes, because this strategy gives LLMs a chance to review and correct the initial answer.
It would be our future work to explore how to make attribute generation better.
\end{tcolorbox}

\subsection{Experiment 3: relationship generation strategy}\label{sec:rel-gen-expr}
\begin{figure}[!tb]
    \centering
    \includegraphics[width=0.9\columnwidth]{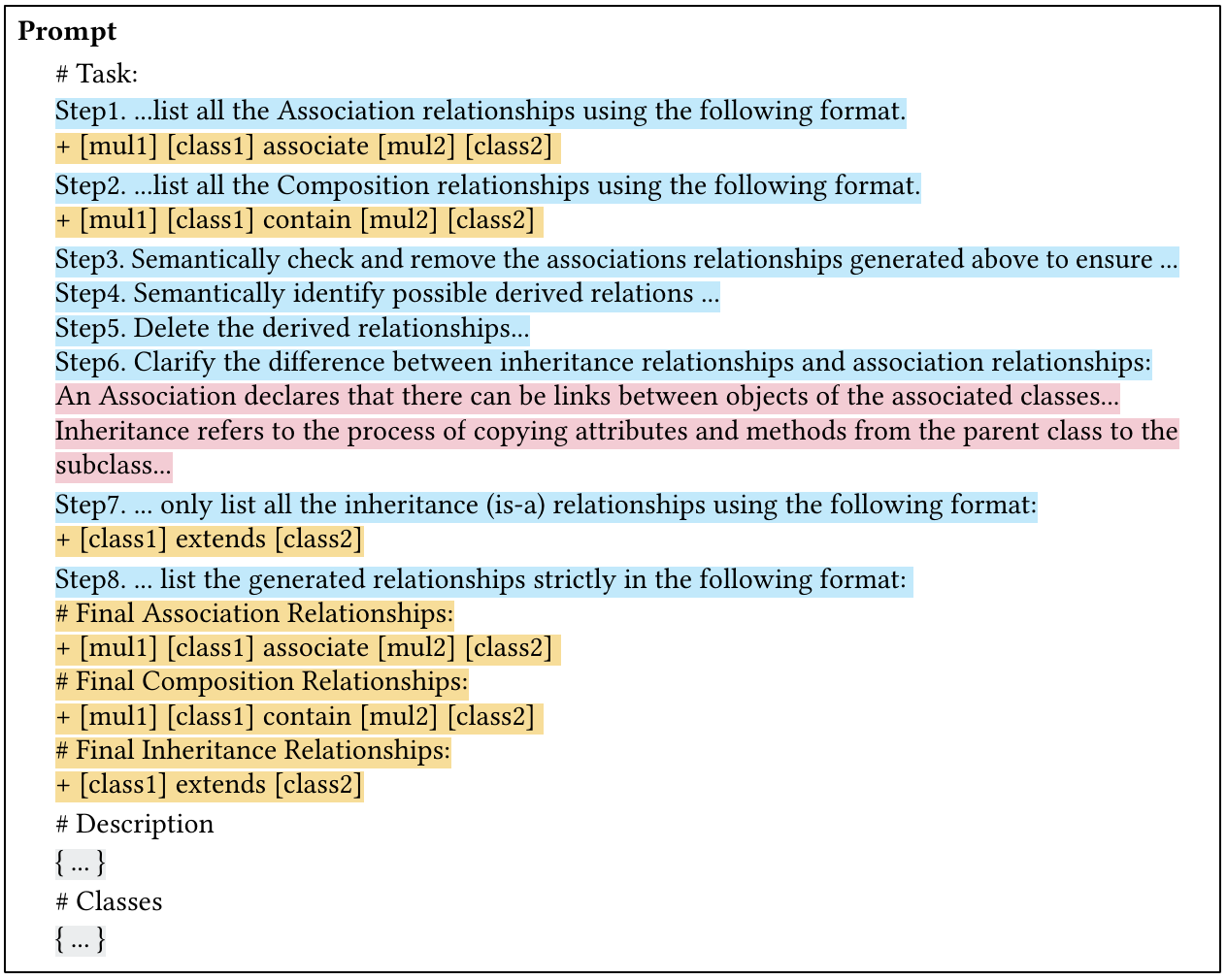}
    \caption{The composite prompt for relationships}
    \label{fig:exp3-single-prompt}
\end{figure}

In Section \ref{sec:relgen}, we think that relationship generation should be divided into two sub-questions, each focusing on a specific relationship type, for complexity management.
The 3rd experiment will evaluate this decomposition strategy by answering RQ3.

\paragraph{Process} We create a prompt (see Figure \ref{fig:exp3-single-prompt}) for relationship generation as a representative of \textit{one-step generation approaches}, by trivially merging the two prompts presented in Section \ref{sec:relgen}.
% We ask our approach and the one-stage approach to generate relationships for \textbf{\textit{DomSys}} and \textbf{\textit{NLPSet}}.
We ask our approach and the one-step approach to generate relationships for all 20 systems.
For each system, both approaches are required to generate 50 models on \gptA{}.
To avoid the influence of classes generation, we use the correct classes extracted from referenced models for relationship generation in this experiment.
By comparing the generated relationships with referenced models, we calculate the average precision, recall, and F1 scores of each approach.
And we choose the temperature 0.9 for associations/aggregations generation, and 0.8 for inheritances generation.

\begin{figure}[!h]
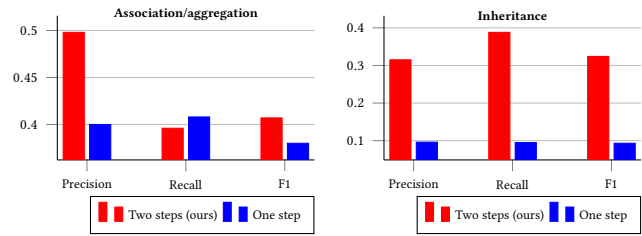

    \centering
    \compbarplot{Association/aggregation}{(1, 0.498) (2, 0.396) (3, 0.407)}{(1, 0.4) (2, 0.408) (3, 0.38)}{Two steps (ours), One step}
    \hspace{0.5em}
    \compbarplot{Inheritance}{(1, 0.315) (2, 0.388) (3, 0.324)}{(1, 0.096) (2, 0.095) (3, 0.093)}{Two steps (ours), One step}
    \caption{Comparison of relationship generation strategies }
    \label{fig:3rd-exp-result}
\end{figure}

\paragraph{Results}
% I need two bar plots for association/aggregation and inheritance. Each bar plot contains 6 bars, denoting prec, recall, f1 of ours and baselines.
Figure \ref{fig:3rd-exp-result} depicts the comparison of precision, recall, and F1 score between our approach and the one-step approach.
\textbf{For associations/aggregations,} the average precision, recall, and F1 score of the one-step approach are around 0.4, 0.41, and 0.38, respectively, while those of our approach are around 0.50, 0.40, and 0.41, respectively.
\textbf{For inheritances,} the average precision, recall, and F1 score of the one-step approach are around 0.10, 0.10, and 0.10, respectively, while those of our approach are around 0.32, 0.39, and 0.32, respectively.
For F1 score, our approach outperforms the one-step one, achieving an average improvement of 7.11\% for association/aggregation, and 248.39\%  for inheritance.

% \paragraph{Answer}

\begin{tcolorbox}
\textbf{\textit{Answer to RQ3.}} According to Figure \ref{fig:3rd-exp-result}, our answer to \textbf{RQ3} is \textbf{YES}---the decomposition of relationship generation significantly improves the results in terms of precision, recall, and F1 score, respectively, compared to the one-step generation.
Our approach breaks down the complex question into smaller ones, enabling LLMs to concentrate on instructions more effectively and solve simple problems.
\end{tcolorbox}

We must emphasize that we input the correct classes into LLMs for relationship generation in this experiment.
This differs from the setup in Experiment 1, which uses the generated classes extracted from Step 2.
Comparing the results of Experiments 1 and 3, we observe that there is a significant causal relation between the accuracy of relationship generation and that of class generation. 
When inputting the correct classes, the generated relationships are more coherent and logical.
However, when inputting the wrong classes, they may mislead LLMs to generate the wrong relationships.
% when inputting the wrong classes, they may lead LLMs to generate the wrong relationships.”
% The accurate generation of classes reduces ambiguity in the model's understanding of relationships.
% enabling the model to more effectively discern subtleties and complexities, thereby yielding more precise results.
% % For association generation, if only one correct class within a relationship is generated, it is challenging for the relationship to be accurately produced. 
% For inheritance relationships, if only the parent or child class is generated, further analysis combining the class and system description is necessary to determine the presence of a child class.

We believe that relationship generation is inherently more complex. 
This is because, in contrast to classes, the expression of a relationship is not necessarily obvious in textual descriptions.
For instance, the sentence ``The selected block follows the mouse movement without overlapping the other blocks and exits the game zone. The selected block can't move near other blocks at least 0.0x from the other blocks.'' states a relationship (i.e., `1 Movement contain 1 Position').
Although this description does not explicitly mention the statement ``1 Movement contain 1 Position'', according to the logic of Movement, we can analyze that the movement contains the change of position.
% However, the description does not explicitly use a ``Position'' object or concept to describe the specific position of the block, but according to the logic of Movement, we can infer that the movement contains the change of position.
This also illustrates that relationship generation requires higher abstraction and system analysis capabilities.
This is a challenge for LLMs.
% It requires not only the identification of previously generated classes but also the localization of relational statements between classes within the system description, and the analysis to determine the type of relationship. 
% This process demands a higher level of semantic understanding and reasoning from the model.

\subsection{Experiment 4: temperatures}\label{sec:temperature-expr}

The temperature is an important parameter that controls the randomness and creativity of LLM's output.
The goal of the 4th experiment is to empirically investigate the impact of temperatures on our approach to model generation by answering RQ4.
% To be concrete, we focus on the following research question:
% \begin{itemize}
%     \item \textbf{RQ4.} \textit{What temperature should be used in our approach?}
% \end{itemize}

\paragraph{Process} For each sub-question of model generation in our approach, i.e. class and attribute generation, association/aggregation generation, and inheritance generation, we test them with different temperature settings, varying from 0.1 to 1.0 with a step size of 0.1.
We perform the tests based on the systems in \textbf{\textit{DomSys}} and \gptA{}.
Given a system in \textbf{\textit{DomSys}}, we conduct 50 tests for each temperature setting.
To evaluate class and attribute generation, we ask \gptA{} to generate classes and attributes based on system descriptions using the prompts in Figure \ref{fig:class-gen}.
To evaluate relationship generation, we ask \gptA{} to generate relationships based on the class names in referenced models, to avoid the influence of class generation, using the prompts in Figures \ref{fig:rel-gen-assoc} and \ref{fig:rel-gen-inherit}.
We compare the results with referenced models, and calculate the average F1 scores under each temperature setting.
Finally, we choose the \textit{best} temperatures that enable the three tasks to achieve their highest F1 scores.
Because we use the chosen temperatures in all the other experiments, we only use \textbf{\textit{DomSys}} and \gptA{} in this experiment to avoid overfitting.

\begin{figure}
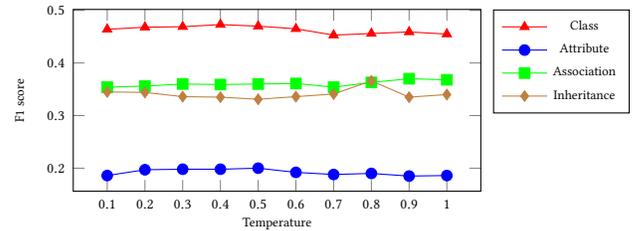

    \centering
    \largetemplot{
        \addlineWithLegend{Class}{triangle*,red}{(0.1,0.464)(0.2,0.468) (0.3,0.469) (0.4,0.473) (0.5,0.470) (0.6,0.465) (0.7,0.453) (0.8,0.456)(0.9,0.459)(1.0,0.455)}
        \addlineWithLegend{Attribute}{*,blue}{(0.1,0.186)(0.2,0.197) (0.3,0.198) (0.4,0.198) (0.5,0.200) (0.6,0.192) (0.7,0.188) (0.8,0.190)(0.9,0.185)(1.0,0.186)}
        \addlineWithLegend{Association}{square*,green}{(0.1,0.354)(0.2,0.356) (0.3,0.36) (0.4,0.359) (0.5,0.360) (0.6,0.361) (0.7,0.354) (0.8,0.363)(0.9,0.37)(1.0,0.368)}
        \addlineWithLegend{Inheritance}{diamond*,brown}{(0.1,0.345)(0.2,0.344) (0.3,0.336) (0.4,0.335) (0.5,0.331) (0.6,0.336) (0.7,0.341) (0.8,0.366)(0.9,0.335)(1.0,0.340)}

    }
    \caption{F1 score variation with temperature changes}\label{fig:4th-exp-overall}
\end{figure}

% all the results should be presented using line charts
\paragraph{Results}
% First, I need a line chart consisting of three lines for classes, assocations, inheritances. X-axis is temperature. Y-axis is f1 score.
% First, I describe the result
Figure \ref{fig:4th-exp-overall} demonstrates how the F1 scores for class generation, attribute generation, association/aggregation generation, and inheritance generation fluctuate across different temperatures.
\begin{itemize} %% FIXME
    \item For class generation and attribute generation, their F1 scores vary from 0.45 to 0.48 and 0.16 to 0.20, respectively. Both generation tasks reach their best results when the temperature is 0.4.
    \item For association/aggregation generation and inheritance generation, their F1 scores change from 0.35 to 0.37 and 0.33 and 0.37, respectively. When the temperature is set to 0.9, association/aggregation generation task achieves the best result. When the temperature is set to 0.8, inheritance generation task achieves the best result. 
\end{itemize}

% Second, for each prompt, draw a single line chart for all the systems. Each system is denoted by a line. (3 line charts in total)

% \begin{itemize} %% FIXME
%     \item For class generation, six systems (i.e., BTMS, H2S, LRMS, DBA, TiOA, and HBMS) obtain their highest F1 scores with high temperatures (i.e., 0.6--0.8), while three systems (i.e., CeOS, TSS, and OTS) need cold temperatures (i.e., 0.2--0.4).
%     \item For attribute generation, six systems (i.e., H2S, CeOS, SHAS, OTS, TiOA, and HBMS) obtain their highest F1 scores with high temperatures, while three systems (i.e., BTMS, LRMS, and TSS) need cold temperatures.
%     \item For association/aggregation generation, the peak points of CeOS, BTMS, H2S, LRMS, SHAS, and OTS occur with cold temperatures (i.e., 0.2--0.4), while those of TSS, HBMS, DBA, and TIOA occur with high temperatures (i.e., 0.6--0.8).
%     \item For inheritance generation, more systems (i.e., TSS, CeOS, H2S, LRMS, HBMS, SHAS, and OTS) reach their peak points with cold temperatures, while only two systems, i.e., HBMS and TiOA, achieve the best F1 score with high temperatures, indicating that inheritance generation favours cold temperatures.
% \end{itemize}

% \paragraph{Answer}

\begin{tcolorbox}
\textbf{\textit{Answer to RQ4.}} According to Figure \ref{fig:4th-exp-overall}, our answer to \textbf{RQ4} is that for class generation and attribute generation, the prompt (i.e., Figure \ref{fig:class-gen}) needs a temperature 0.4, and for association/aggregation generation and inheritance generation, the prompts (i.e., Figure's \ref{fig:rel-gen-assoc} and \ref{fig:rel-gen-inherit}) need a temperature 0.8. 
\end{tcolorbox}
% \begin{remark}
% Figure \ref{fig:4th-exp-detail} shows the F1 scores for class, attribute, association, and inheritance generation for each system at every temperature.
% The results show that the optimal temperature is different for different tasks for different systems.
% But combined with Figure \ref{fig:4th-exp-overall}, all systems achieve the best average scores when the temperatures are 0.4 (class and attribute), 0.9 (association), and 0.8 (inheritance), respectively.

In our approach, different generation sub-tasks need different temperatures.
We think that different problems pose different demands for the creativity of LLMs.
Class generation needs more creativity because LLMs must identify the (implicit) classes that exist in the problem domain but are not emphasized in the system description.
However, relationship generation must be based on the description to reflect the system responsibilities.
% \end{remark}

The 4th experiment results are only applicable for \gptA{}.
Different temperature settings may be required for different LLMs.
The temperature setting of this experiment may not generalize to other LLMs, while this experimental method is applicable to other LLMs.

\begin{figure}[!tb]
    \centering
\templotWithLabel{Class}{Temperature}{F1 score}{
\addline{Class}{triangle*,green,mark size=1.5pt,solid,mark options={solid}}{(0.1,0.301)(0.2,0.334)(0.3,0.422)(0.4,0.404)(0.5,0.4)(0.6,0.411)(0.7,0.349)(0.8,0.345)(0.9,0.407)(1.0,0.377)}
\addline{Class}{triangle*,red,mark size=1.5pt,densely dashdotdotted,mark options={solid}}{(0.1,0.506)(0.2,0.54)(0.3,0.525)(0.4,0.523)(0.5,0.521)(0.6,0.51)(0.7,0.507)(0.8,0.528)(0.9,0.498)(1.0,0.5)}
\addline{Class}{diamond*,gray,mark size=1.5pt, densely dashdotted,mark options={solid}}{(0.1,0.474)(0.2,0.454)(0.3,0.458)(0.4,0.445)(0.5,0.447)(0.6,0.419)(0.7,0.408)(0.8,0.432)(0.9,0.405)(1.0,0.413)}
\addline{Class}{square*,blue,mark size=1pt,densely dotted,mark options={solid}}{(0.1,0.374)(0.2,0.371)(0.3,0.364)(0.4,0.376)(0.5,0.375)(0.6,0.379)(0.7,0.383)(0.8,0.394)(0.9,0.412)(1.0,0.396)}
\addline{Class}{+,orange,mark size=1.5pt,solid,mark options={solid}}{(0.1,0.445)(0.2,0.481)(0.3,0.463)(0.4,0.497)(0.5,0.491)(0.6,0.5)(0.7,0.468)(0.8,0.476)(0.9,0.477)(1.0,0.473)}
\addline{Class}{x,cyan,mark size=1.5pt,solid,mark options={solid}}{(0.1,0.48)(0.2,0.534)(0.3,0.536)(0.4,0.534)(0.5,0.53)(0.6,0.525)(0.7,0.511)(0.8,0.514)(0.9,0.472)(1.0,0.469)}
\addline{Class}{pentagon*,violet,mark size=1.5pt,solid,mark options={solid}}{(0.1,0.536)(0.2,0.465)(0.3,0.457)(0.4,0.473)(0.5,0.448)(0.6,0.457)(0.7,0.457)(0.8,0.442)(0.9,0.487)(1.0,0.477)}
\addline{Class}{square*,magenta,mark size=1pt,densely dotted,mark options={solid}}{(0.1,0.259)(0.2,0.255)(0.3,0.25)(0.4,0.275)(0.5,0.28)(0.6,0.25)(0.7,0.272)(0.8,0.24)(0.9,0.266)(1.0,0.269)}
\addline{Class}{pentagon*,brown,mark size=1pt,densely dashdotted,mark options={solid}}{(0.1,0.588)(0.2,0.574)(0.3,0.558)(0.4,0.545)(0.5,0.543)(0.6,0.544)(0.7,0.546)(0.8,0.551)(0.9,0.532)(1.0,0.54)}
\addline{Class}{diamond*,black,mark size=1.5pt,densely dashdotted,mark options={solid}}{(0.1,0.672)(0.2,0.67)(0.3,0.661)(0.4,0.654)(0.5,0.664)(0.6,0.655)(0.7,0.628)(0.8,0.641)(0.9,0.631)(1.0,0.632)}

}
\templotWithLabel{Attribute}{Temperature}{F1 score}{\addline{Class}{triangle*,green,mark size=1.5pt,solid,mark options={solid}}{(0.1,0.3)(0.2,0.315)(0.3,0.356)(0.4,0.357)(0.5,0.354)(0.6,0.361)(0.7,0.34)(0.8,0.331)(0.9,0.351)(1.0,0.335)}
\addline{Class}{triangle*,red,mark size=1.5pt,densely dashdotdotted,mark options={solid}}{(0.1,0.121)(0.2,0.193)(0.3,0.143)(0.4,0.138)(0.5,0.18)(0.6,0.15)(0.7,0.137)(0.8,0.17)(0.9,0.164)(1.0,0.174)}
\addline{Class}{diamond*,gray,mark size=1.5pt, densely dashdotted,mark options={solid}}{(0.1,0.143)(0.2,0.144)(0.3,0.154)(0.4,0.143)(0.5,0.141)(0.6,0.126)(0.7,0.141)(0.8,0.129)(0.9,0.12)(1.0,0.123)}
\addline{Class}{square*,blue,mark size=1pt,densely dotted,mark options={solid}}{(0.1,0.225)(0.2,0.234)(0.3,0.247)(0.4,0.245)(0.5,0.246)(0.6,0.227)(0.7,0.211)(0.8,0.23)(0.9,0.208)(1.0,0.187)}
\addline{Class}{+,orange,mark size=1.5pt,solid,mark options={solid}}{(0.1,0.147)(0.2,0.149)(0.3,0.147)(0.4,0.153)(0.5,0.144)(0.6,0.147)(0.7,0.146)(0.8,0.144)(0.9,0.144)(1.0,0.149)}
\addline{Class}{x,cyan,mark size=1.5pt,solid,mark options={solid}}{(0.1,0.072)(0.2,0.105)(0.3,0.105)(0.4,0.112)(0.5,0.109)(0.6,0.114)(0.7,0.109)(0.8,0.119)(0.9,0.082)(1.0,0.101)}
\addline{Class}{pentagon*,violet,mark size=1.5pt,solid,mark options={solid}}{(0.1,0.355)(0.2,0.353)(0.3,0.349)(0.4,0.359)(0.5,0.337)(0.6,0.336)(0.7,0.325)(0.8,0.322)(0.9,0.316)(1.0,0.307)}
\addline{Class}{square*,magenta,mark size=1pt,densely dotted,mark options={solid}}{(0.1,0.063)(0.2,0.054)(0.3,0.049)(0.4,0.054)(0.5,0.052)(0.6,0.049)(0.7,0.053)(0.8,0.043)(0.9,0.057)(1.0,0.066)}
\addline{Class}{pentagon*,brown,mark size=1pt,densely dashdotted,mark options={solid}}{(0.1,0.165)(0.2,0.162)(0.3,0.165)(0.4,0.164)(0.5,0.168)(0.6,0.156)(0.7,0.166)(0.8,0.164)(0.9,0.157)(1.0,0.164)}
\addline{Class}{diamond*,black,mark size=1.5pt,densely dashdotted,mark options={solid}}{(0.1,0.269)(0.2,0.264)(0.3,0.266)(0.4,0.256)(0.5,0.266)(0.6,0.258)(0.7,0.25)(0.8,0.249)(0.9,0.255)(1.0,0.252)}

}
\templotWithLabel{Association}{Temperature}{F1 score}{
\addline{Class}{triangle*,green,mark size=1.5pt,solid,mark options={solid}}{(0.1,0.467)(0.2,0.464)(0.3,0.49)(0.4,0.478)(0.5,0.472)(0.6,0.475)(0.7,0.47)(0.8,0.475)(0.9,0.483)(1.0,0.488)}
\addline{Class}{triangle*,red,mark size=1.5pt,densely dashdotdotted,mark options={solid}}{(0.1,0.301)(0.2,0.271)(0.3,0.301)(0.4,0.294)(0.5,0.291)(0.6,0.296)(0.7,0.304)(0.8,0.305)(0.9,0.327)(1.0,0.322)}
\addline{Class}{diamond*,gray,mark size=1.5pt, densely dashdotted,mark options={solid}}{(0.1,0.402)(0.2,0.48)(0.3,0.466)(0.4,0.433)(0.5,0.427)(0.6,0.447)(0.7,0.416)(0.8,0.407)(0.9,0.422)(1.0,0.407)}
\addline{Class}{square*,blue,mark size=1pt,densely dotted,mark options={solid}}{(0.1,0.458)(0.2,0.463)(0.3,0.456)(0.4,0.458)(0.5,0.47)(0.6,0.489)(0.7,0.483)(0.8,0.473)(0.9,0.496)(1.0,0.463)}
\addline{Class}{+,orange,mark size=1.5pt,solid,mark options={solid}}{(0.1,0.507)(0.2,0.541)(0.3,0.518)(0.4,0.535)(0.5,0.501)(0.6,0.522)(0.7,0.519)(0.8,0.542)(0.9,0.534)(1.0,0.51)}
\addline{Class}{x,cyan,mark size=1.5pt,solid,mark options={solid}}{(0.1,0.332)(0.2,0.28)(0.3,0.294)(0.4,0.286)(0.5,0.309)(0.6,0.303)(0.7,0.296)(0.8,0.299)(0.9,0.314)(1.0,0.325)}
\addline{Class}{pentagon*,violet,mark size=1.5pt,solid,mark options={solid}}{(0.1,0.327)(0.2,0.244)(0.3,0.255)(0.4,0.284)(0.5,0.297)(0.6,0.3)(0.7,0.298)(0.8,0.317)(0.9,0.342)(1.0,0.347)}
\addline{Class}{square*,magenta,mark size=1pt,densely dotted,mark options={solid}}{(0.1,0.234)(0.2,0.281)(0.3,0.27)(0.4,0.282)(0.5,0.293)(0.6,0.251)(0.7,0.246)(0.8,0.282)(0.9,0.256)(1.0,0.281)}
\addline{Class}{pentagon*,brown,mark size=1pt,densely dashdotted,mark options={solid}}{(0.1,0.328)(0.2,0.372)(0.3,0.388)(0.4,0.371)(0.5,0.357)(0.6,0.35)(0.7,0.314)(0.8,0.33)(0.9,0.311)(1.0,0.332)}
\addline{Class}{diamond*,black,mark size=1.5pt,densely dashdotted,mark options={solid}}{(0.1,0.187)(0.2,0.161)(0.3,0.157)(0.4,0.169)(0.5,0.181)(0.6,0.178)(0.7,0.194)(0.8,0.198)(0.9,0.215)(1.0,0.206)}

}
\templotWithLabel{Inheritance}{Temperature}{F1 score}{
\addline{Class}{triangle*,green,mark size=1.5pt,solid,mark options={solid}}{(0.1,0.0)(0.2,0.0)(0.3,0.0)(0.4,0.0)(0.5,0.0)(0.6,0.0)(0.7,0.0)(0.8,0.0)(0.9,0.0)(1.0,0.0)}
\addline{Class}{triangle*,red,mark size=1.5pt,densely dashdotdotted,mark options={solid}}{(0.1,0.85)(0.2,0.857)(0.3,0.857)(0.4,0.857)(0.5,0.85)(0.6,0.833766)(0.7,0.835195)(0.8,0.857)(0.9,0.841428571)(1.0,0.808)}
\addline{Class}{diamond*,gray,mark size=1.5pt, densely dashdotted,mark options={solid}}{(0.1,0.0)(0.2,0.0)(0.3,0.0)(0.4,0.0)(0.5,0.0)(0.6,0.0)(0.7,0.0)(0.8,0.0)(0.9,0.0)(1.0,0.0)}
\addline{Class}{square*,blue,mark size=1pt,densely dotted,mark options={solid}}{(0.1,0.684358)(0.2,0.727)(0.3,0.618)(0.4,0.596)(0.5,0.63)(0.6,0.65557)(0.7,0.659711)(0.8,0.96)(0.9,0.709242247)(1.0,0.77)}
\addline{Class}{+,orange,mark size=1.5pt,solid,mark options={solid}}{(0.1,0.0)(0.2,0.0)(0.3,0.0)(0.4,0.0)(0.5,0.0)(0.6,0.0)(0.7,0.0)(0.8,0.0)(0.9,0.0)(1.0,0.0)}
\addline{Class}{x,cyan,mark size=1.5pt,solid,mark options={solid}}{(0.1,0.373085)(0.2,0.329)(0.3,0.385)(0.4,0.398)(0.5,0.342)(0.6,0.378095)(0.7,0.368671)(0.8,0.37)(0.9,0.367421356)(1.0,0.346)}
\addline{Class}{pentagon*,violet,mark size=1.5pt,solid,mark options={solid}}{(0.1,0.361592)(0.2,0.393)(0.3,0.339)(0.4,0.382)(0.5,0.368)(0.6,0.378125)(0.7,0.394625)(0.8,0.33)(0.9,0.367062785)(1.0,0.379)}
\addline{Class}{square*,magenta,mark size=1pt,densely dotted,mark options={solid}}{(0.1,0.672195)(0.2,0.655)(0.3,0.685)(0.4,0.613)(0.5,0.679)(0.6,0.676573)(0.7,0.659776)(0.8,0.726)(0.9,0.581362859)(1.0,0.604)}
\addline{Class}{pentagon*,brown,mark size=1pt,densely dashdotted,mark options={solid}}{(0.1,0.447403)(0.2,0.47)(0.3,0.465)(0.4,0.47)(0.5,0.427)(0.6,0.417752)(0.7,0.43004)(0.8,0.416)(0.9,0.404741703)(1.0,0.438)}
\addline{Class}{diamond*,black,mark size=1.5pt,densely dashdotted,mark options={solid}}{(0.1,0.060217)(0.2,0.007)(0.3,0.011)(0.4,0.037)(0.5,0.014)(0.6,0.017231)(0.7,0.058263)(0.8,0.0)(0.9,0.083434343)(1.0,0.054)}

}
\caption{The variation of F1 score for all systems (each line denotes a certain system: 
    \includegraphics[height=8pt]{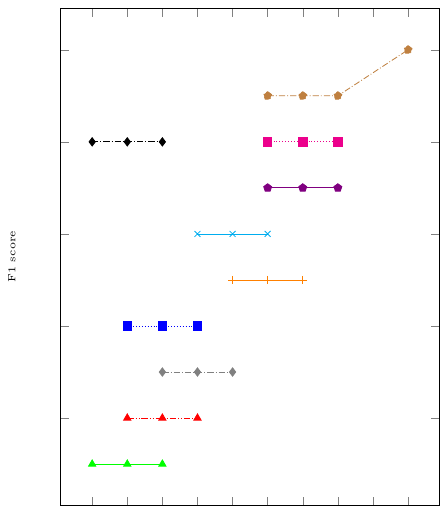} for BTMS,
    \includegraphics[height=8pt]{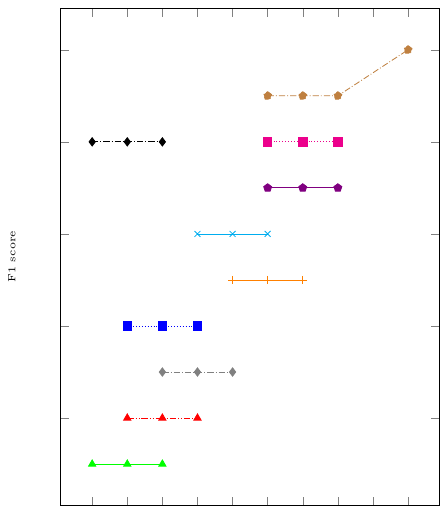} for H2S,
    \includegraphics[height=8pt]{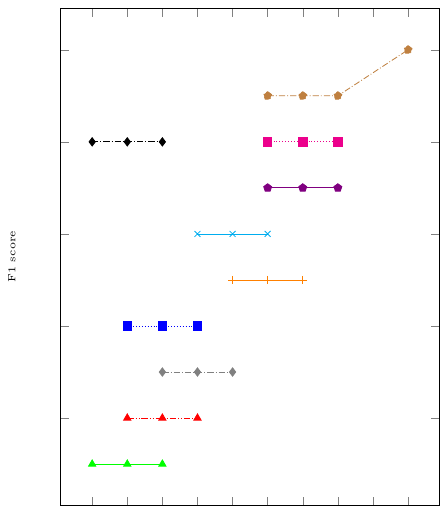} for LRMS,
    \includegraphics[height=8pt]{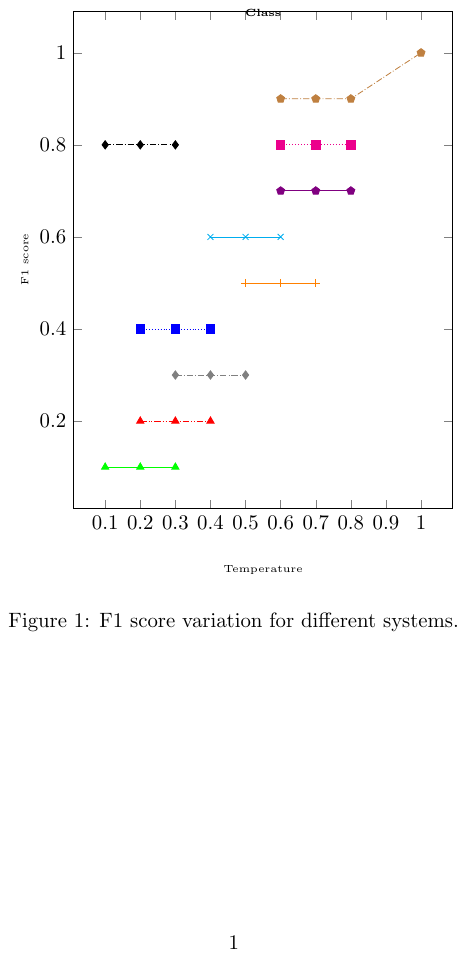} for CeOS,
    \includegraphics[height=8pt]{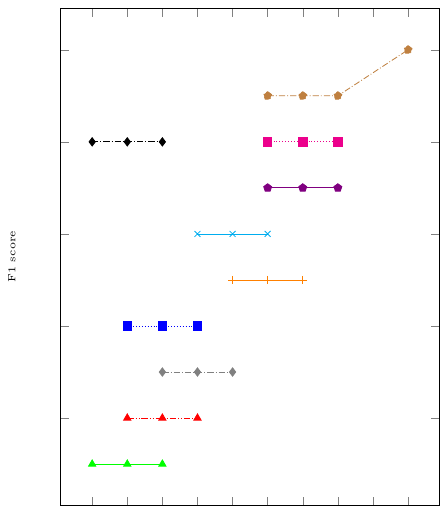} for TSS,
    \includegraphics[height=8pt]{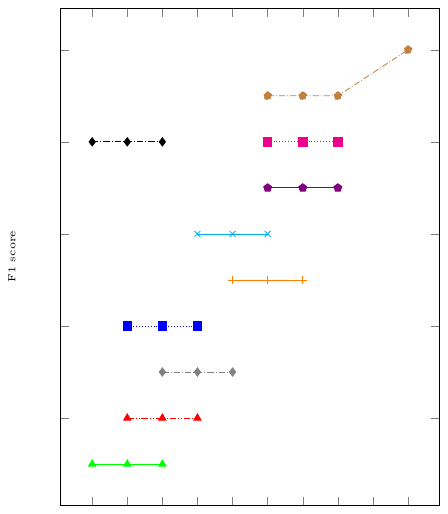} for SHAS,
    \includegraphics[height=8pt]{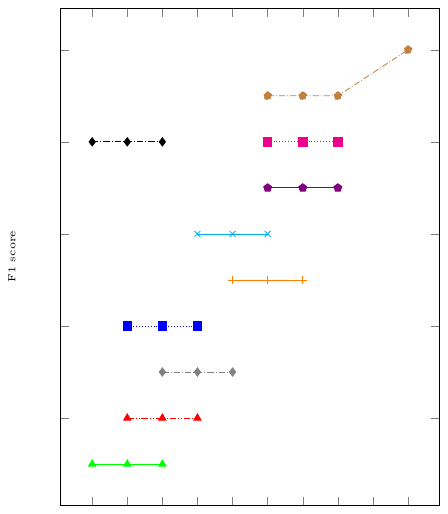} for OTS,
    \includegraphics[height=8pt]{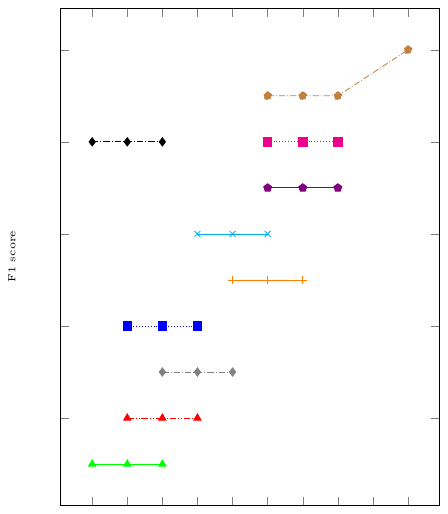} for DBA,
    \includegraphics[height=8pt]{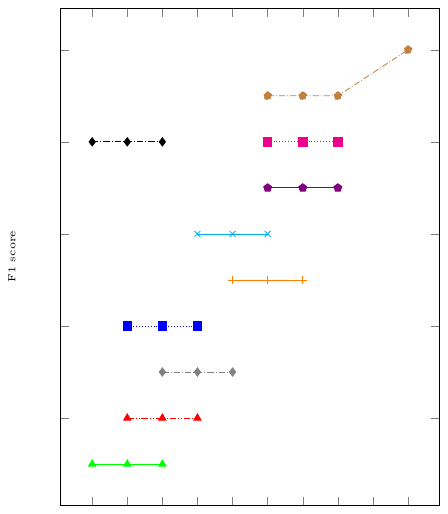} for TiOA,
    and \includegraphics[height=8pt]{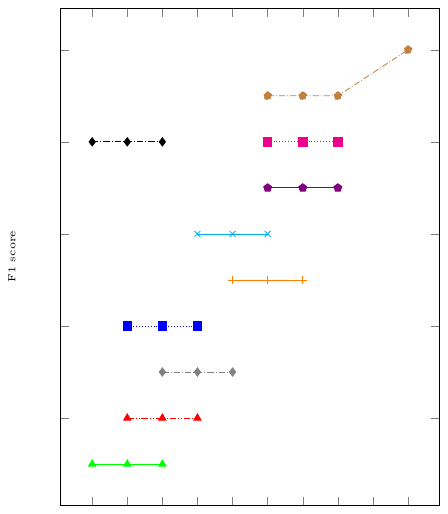} for HBMS
    )}
    \label{fig:4th-exp-detail}
\end{figure}

According to Figures \ref{fig:4th-exp-detail}, different systems are sensitive to different temperatures.
For classes, some systems (e.g., LRMS, CeOS) have higher scores at low temperatures (< 0.5), while some systems (e.g., TiOA) have higher scores at high temperatures (> 0.5).
For associations, some systems (e.g., CeOS, SHAS) have higher scores at low temperatures (< 0.5), while some systems (e.g., LRMS, TSS, OTS) have higher scores at high temperatures (> 0.5).
It would be \textit{an interesting future work to investigate the causality between the temperature and the characteristics of the system} (i.e., domain and complexity).
%\subsection{Case explorations}

%% file: section/threats.tex
\section{Threats to validity}
We mainly focus on the external validity, the construct validity, and the conclusion validity of this study.
Because this paper does not investigate a casual relationship, we ignore the internal validity.

\textbf{External validity} refers to whether the evaluation results can be generalized to other contexts.
We only evaluate our approach with 20 systems.
It is possible that our approach might not be effective for other systems.
To mitigate this issue, we should try more systems and LLMs.
However, due to the lack of benchmarks for model generation, we are currently unable to conduct a comprehensive study.
Another threat is that we test our approach on \gptA{} and \llama{} only.
The results might vary on other LLMs.
However, \gptA{} is one of the most popular LLMs and can provide its users with a good balance of capability and cost.
We assume that the results of our study are still valuable at the current stage.
It would be our future work to try different LLMs.

The randomness of LLMs (especially, public LLMs) is also a threat to the external validity.
The answers obtained by querying an LLM at different times may have different quality.
To mitigate this issue, we ran each test 50 times to gain the average result, since the result usually converged after 20 tests.
It would be our future work to enhance our evaluation of a locally deployed LLM.

Different temperature settings may be different for different LLMs. 
% The temperature setting is required for different LLMs.
Experiments 1, 2, and 3 all utilized the temperature settings from Experiment 4, which were based on \gptA{}, specifically 0.4 (classes), 0.9 (associations/aggregations), and 0.8 (inheritances). 
These temperatures may not apply to other LLMs.
% For other LLMs, the temperatures need to be re-evaluated according to the methodology outlined in Experiment 4.
% Then all other experiment settings can follow other experimental process based the temperatures.
For other LLMs, the temperatures need to be re-evaluated according to the process outlined in Experiment 4. 
Subsequently, all other experiments can follow the experimental process in this paper based on these temperatures.

\textbf{Construct validity} refers to the degree to which the metrics used in a study measure the performance of our approach.
In this section, we mainly use precision, recall, and F1 scores that are calculated by matching generated models with Oracle models, following pioneering work.
Precision, recall, and F1 scores are frequently used in similar contexts involving prediction, so they are reasonable metrics in this study.
There might be better metrics for model generation, just like CodeBLEU \cite{ren2020codebleu} and \textit{Pass@K} for source code.
However, to the best of our knowledge, we are unaware of such metrics in the MDE community.
It would be our future work to design better metrics for the automatic evaluation of model generation.

% Reliablity
\textbf{Conclusion validity} refers to the reliability and accuracy of the conclusions drawn from a study.
To avoid a threat to conclusion validity, all results and answers to the research questions in our evaluation were discussed until the authors reached an agreement.

%% file: section/conclusion.tex
\section{Conclusion}\label{sec:conclusion}
% 我们提出了一个领域建模的问题分解方案，旨在以分而治之提高大模型解决问题的能力。
% 我们研究llm在促进软件开发过程中的领域建模方面的潜力。
% 具体来说，基于问题分解思想和面向对象建模方法，我们将建模任务分解为了：类和属性生成，关联和聚合生成，继承生成，旨在分布每个子任务的复杂性。
% 基于20个不同系统的评估表明，我们的方法证明了方案的有效性，在recall precision值比单个提示方法表现出色。
% 并且我们也开发了原型工具总结最后的建模结果，实现从建议到可执行操作的无缝过渡。
% 我们可以通过案例研究展示了LLM在现实世界领域建模中的潜力。It can be a starting point for a discussion on LLM scalability
% 此外，我们期望我们的工作能够帮助开发工具和技术，特别是基于llm的，为领域建模而设计的工具和技术。 We expect our work to aid development of tools and techniques, especially LLM-based, designed for domain modeling.

% conclusion
This paper proposes an LLM-based domain modeling approach by adopting the idea of question decomposition.
Following modeling guidelines, we divide the problem of model generation into three sub-problems of class generations, association generation, and inheritance generation, to reduce the complexity of each sub-problem.
By combining AI with human knowledge, we have carefully designed the prompts and the workflow of our approach.
The evaluation based on 20 systems demonstrates on GPT-3.5-Turbo and Llama-3-8B that our approach outperforms the approach proposed by pioneering work, which is based on a single prompt.

% limitation 
At current stage, we are aware that our approach has many limitations.
For example, our approach cannot predict the names of associations (aggregations), and the overall F1 score should be further improved before it can be put into practice.
% Our method did not do relevant experiments based on gpt4. 
% This is because the temperature settings of different models may be different mentioned above, and this is one of the working points we will explore in the future.
With the advent of GPT-4, it would be our future work.
Nevertheless, we believe that this paper represents a step forward in addressing AI-based model generation.
We expect that the design rationales of our approach and the findings of our study may facilitate further discussions and improvements on this topic.

% future work
Regarding the future work, we first plan to refine our approach, especially by adopting more advanced prompt engineering techniques.
Second, we will enhance our prototype tool in functionality and usability, and will contribute it to the MDE community.
Third, we plan to improve our evaluation by testing more systems on different LLMs.
Finally, we plan to explore the potential for fine-tuning LLMs for model generation.

Along with the paper, we have also submitted a data package containing the raw experimental data and Python scripts for review.
Our tool introduced in Section \ref{sec:tool} can be provided upon request.